\documentclass[journal,comsoc]{IEEEtran}

\usepackage{graphicx}
\usepackage{amsmath}
\usepackage{amssymb}
\usepackage[noadjust]{cite}
\usepackage{float}
\usepackage{algorithm}
\usepackage{bibentry}
\usepackage{balance}
\usepackage{algorithm}
\usepackage{algorithmicx}
\usepackage{algpseudocode}
\usepackage{xcolor}
\usepackage{hyperref}
\newtheorem{proposition}{Proposition}
\graphicspath{{img/}}

\begin{document}

\title{ Dynamic Interference Management for  UAV-Assisted Wireless Networks}
	
\author{Ali Rahmati,~\IEEEmembership{Student Member,~IEEE,}  Seyyedali Hosseinalipour, \IEEEmembership{Student Member,~IEEE,} Yavuz Yapici, \IEEEmembership{Member,~IEEE,} Xiaofan He, \IEEEmembership{Member,~IEEE,}  Ismail Guvenc, \IEEEmembership{Senior Member,~IEEE,} Huaiyu Dai, \IEEEmembership{Fellow,~IEEE,} and Arupjyoti~Bhuyan, \IEEEmembership{Senior Member,~IEEE}
\thanks{Part of the material will be presented in IEEE Globecom 2019 \cite{rahmati2019interference}.}
\thanks{
A. Rahmati, S. Hosseinalipour, Y. Yapici, \.{I}. G\"{u}ven\c{c}, and H. Dai are with the Department of Electrical and Computer Engineering, North Carolina State University, Raleigh, NC (e-mail:~\{arahmat, shossei3, yyapici, iguvenc, hdai\}@ncsu.edu).}
\thanks{X. He is with School of Electronic Information, Wuhan University, Wuhan, China (e-mail: xiaofanhe@whu.edu.cn). }
\thanks{A. Bhuyan is with Idaho National Laboratory, Idaho Falls, ID (e-mail:~arupjyoti.bhuyan@inl.gov). }
\thanks{This work is supported in part through the INL Laboratory Directed Research \& Development (LDRD) Program under DOE Idaho Operations Office Contract DE-AC07-05ID14517.}}


\maketitle

\vspace{-14mm}
\begin{abstract}
The deployment of unmanned aerial vehicles (UAVs) is proliferating as they are effective, flexible and cost-efficient  devices for a variety of applications ranging from natural disaster recovery to delivery of goods. We investigate a transmission mechanism aiming to improve the data rate between a  base station (BS) and a user equipment through deploying multiple relaying UAVs. We consider the effect of interference, which is incurred by the  nodes of another established communication network. Our primary goal is to design the 3D trajectories and power allocation for the  UAVs to maximize the data flow while  the interference constraint is met.  The UAVs can reconfigure their locations to evade  the unintended/intended interference caused by  reckless/smart interferers. We also  consider the scenario in which   smart jammers chase the UAVs  to degrade the communication  quality. In this case, we investigate the problem from the perspective of both UAV network and smart jammers. An alternating-maximization approach is proposed to address the joint 3D trajectory design and power allocation problem.  We handle the 3D trajectory design  by resorting to spectral graph theory and subsequently address the power allocation through convex optimization techniques. Finally, we demonstrate the efficacy of our proposed method through
simulations.
\end{abstract}


\section{Introduction}
\noindent \IEEEPARstart{T}{he}  utilization of unmanned aerial vehicles (UAVs) has recently become a practical approach for a variety of mission-driven applications including border surveillance, natural disaster aftermath, monitoring, search and rescue, and purchase delivery \cite{hayat2016survey,rahmati2019energy,ERDELJ201772,8255734,rabta2018drone, 8812925}. Owing to the low acquisition cost of UAVs as well as their fast deployment and efficient coverage capabilities, UAV-assisted wireless communications has attracted extensive interest recently \cite{mag,7577063,mag2,app5G1,mozaffari2019beyond,8469055, kumbhar2018interference}. Specifically, the 3D mobility feature of UAVs and the coexistence of relaying UAVs with other existing communication networks (e.g., cellular networks) have led to new design challenges and opportunities in these networks~\cite{8470897,8316776}, {\color{black}which are not investigated in the context of classic wireless sensor networks~\cite{YOUNIS2008621}}.  This fact has promoted an extensive literature dedicated to studying the unique design aspects of these networks, e.g.,~\cite{zhan2006wireless,wang2017improving,Channel,8424236,hosseinalipour2019interference,faqir2018energy,8116613, zhang2018joint, zeng2016throughput,ono2016wireless ,chen2018local}.
In current literature, the UAV-assisted relay communication is mainly studied in two different contexts, in which the  network is assumed to be either \textit{static}, i.e., the positions of the UAVs, transmitter, and receiver are fixed during the data transmission~\cite{chowdhury2019effects, zhan2006wireless,wang2017improving, Channel,8424236, hosseinalipour2019interference}, or \textit{dynamic}~\cite{faqir2018energy,8116613, zhang2018joint,zeng2016throughput, ono2016wireless,chen2018local}, where the positions of the transmitter and receiver are assumed to be fixed while the UAVs are typically assumed to be mobile.

In the context of static UAV-assisted wireless communications, in \cite{chowdhury2019effects}, considering an interference limited in band downlink cellular network, the authors studied the effects of scheduling criteria, mobility constraints, path loss models, backhaul
constraints, and 3D antenna radiation pattern on trajectory
optimization problem of an UAV. In \cite{zhan2006wireless}, optimal deployment of a
UAV in a wireless relay communication
system is obtained in order to improve the quality of communications between
two obstructed access points by maximizing the average data rate of the system, while limiting the symbol
error rate below a threshold.
In \cite{wang2017improving}, a relay network is considered in the context of a four node channel setup consisting of a transmitter, a receiver, a UAV relay, and an eavesdropper, where the goal is to shed  light on the application of UAV-enabled relaying in secure
wireless communications. The secrecy rate maximization problem is formulated, which turns out to be non-convex, for which an iterative approach based on difference of concave (DC) programming is proposed.
 In~\cite{Channel}, the UAVs are utilized to form an aerial backhaul network so as to enhance the performance of the ground network, which is measured through data rate and delay. The link configuration between the UAVs and the gateways, and among the UAVs, is formulated as a network formation game, which is solved through a myopic network formation algorithm. 
 Considering multiple static UAVs, optimal UAV locations are derived in \cite{8424236} through maximizing the data rate in single link multi-hop and multiple links dual-hop relaying schemes. In this work, it is assumed that the UAVs are hovering at an identical fixed altitude during the transmission. As a follow-up work for \cite{8424236},
 in our recent work~\cite{hosseinalipour2019interference, hosseinalipour2019interference2}, we studied the optimal position planning of UAV relays  between the transmitter and the receiver, which coexist with a major interferer in the environment. In this work, we investigated the following two new problems: i) identifying the minimum required number of UAVs and their optimal positions to satisfy a given SIR of the system, ii) developing a distributed algorithm to maximize the SIR of the system requiring message exchange only between adjacent UAV relays.

 In the context of \textit{dynamic} UAV-assisted wireless communication, in \cite{faqir2018energy}, the joint optimization of propulsion and transmission energies for a UAV relay-assisted communication network is studied. A general optimal control problem is formulated for energy minimization based on dynamic models for both transmission and mobility.
 In \cite{8116613}, the optimum altitude of a UAV for
both static and mobile relaying, which corresponds to the circular movements around the user, is considered so as to maximize the reliability of the system, which is measured through total power loss,
 the overall outage, and the overall bit error rate. It is shown that 
that decode-and-forward relaying is better than amplify-and-forward relaying in terms of reliability. In \cite{zhang2018joint}, a UAV-assisted relay communication network is proposed, where the UAV serves as a relay between a base station and a mobile device. The amplify-and-forward relaying scheme is used, for which the trajectory of UAV,
the transmit power of both the UAV and the mobile device are obtained  so as to minimize
the outage probability of the system.
 In \cite{zeng2016throughput}, assuming a source-destination pair and a UAV relay,  an end-to-end throughput
maximization problem  is formulated to optimize the
relay trajectory and the source/relay power allocations subject to practical constraints on the UAV speed, transmitting power of the transmitter and the UAV, etc. Afterward, an alternating optimization approach is proposed to jointly derive the optimal transmission power of the transmitter and the UAV, and the UAV trajectory. 
In \cite{ono2016wireless},  a wireless relay network
model is considered, in which a fixed-wing UAV 
serves as a  relay among the ground stations
with disconnected communication links in the event of disasters. It is assumed that the UAV deploys the decode-and-forward relaying protocol. Considering the fact that in contrast to rotatory-wing UAVs, fixed-wing UAVs require circular movements to maintain their altitude, it is shown that the conventional fixed rate relaying will  no longer be effective. To this end, a variable rate relaying approach is proposed to enhance the performance of the system measured through outage probability and information rate.
In \cite{chen2018local}, UAVs are deployed in a wireless network in order to provide connectivity or boost the capacity for the ground users. A nested segmented propagation model is proposed for the air-to-ground channel, based on which they proposed an algorithm to search  the optimal UAV position for establishing the best wireless relay link between a base station and a user in a dense urban area. 

In this work, we consider the  application of UAVs in a more complicated relay network structure, where the network consists of multiple ground/terrestrial nodes and aerial nodes, i.e., UAVs. The direction for the flow of information is assumed to be time varying and thus unknown a priori. In this context, each node is considered to be a transceiver.  We aim to propose an analytical  framework to enhance the current literature on the subject  by incorporating the existence of interferers. We consider the mobility of UAVs during the data transmission to avoid/suppress the interference from  multiple interferers. 

The existence of multiple interferers  in a 3D environment makes our methodology different from the current literature. We pursue two different design schemes considering two different interpretations for the interferers: (\textit{i}) Reckless interferer and (\textit{ii}) Smart interferer. Reckless interferers produce  interference unintentionally, e.g., primary transmitters in the context of cognitive radio networks. In this case, we study the problem considering two scenarios; First, we assume that  the UAV network and the  co-existing network (which is referred to as primary network) can  cooperate  to mitigate the mutual interference; Second, these two networks are assumed to be non-cooperative.   We consider both the interference from the primary network to the UAV network and vice versa as the satisfactory performance of both networks is of high importance. Smart interferers, on the other hand, generate interference by intention to interrupt or degrade the communication quality of the UAV network, e.g., mobile jammers, or malicious UAV users. Here, we consider the problem from the perspective of both the UAV network and the smart jammers.  In this case, we only consider the interference from the smart jammers to the UAV network. We pursue the problem of joint power allocation  and 3D trajectory design for UAVs to maximize the achievable data rate of the network.
 
 \subsection{Summary of Contributions}
 \begin{itemize}
    \item We investigate the UAV-assisted communication problem   in the presence of interference based on graph theory. In particular, we formulate the optimization problem as a single commodity maximum flow problem. Considering an interference limited scenario, we deploy a modified signal to interference ratio (SIR) so as to guarantee the safety separation of the UAVs in the 3D trajectory design. We assume different interpretations for the interferers as (\textit{i}) Reckless interferer (e.g.  transmitters in the primary network) and (\textit{ii}) Smart interferer (e.g. jammers).

  \item Assuming a reckless interferer, we propose a  3D trajectory design and power allocation scheme in order to improve the transmission flow in the UAV network while an interference constraint is met in the primary network. The 3D trajectory design is addressed using spectral graph theory in which the UAVs try to reconfigure their locations in order to evade the unwanted interference. The power allocation design is done using successive convex approximation (SCA) approach to make sure that the interference constraint on the primary network is satisfied.

     \item  In  the case of smart interferers, we approach the problem from both the UAV network and the  smart interferers' perspectives. We propose a 3D trajectory design for  the legitimate UAVs in the relay network and the smart interferer. The novel idea for the UAVs in the relay network is to move in a direction  to evade  the interference caused by the smart interferers. On the other hand, from the smart interferer's perspective, the smart UAVs  aim to chase the legitimate UAVs in the relay network so as to increase the intended interference which leads to degradation of the communication flow between the BS to the user equipment (UE).


\end{itemize}
 
 The rest of the paper is organized as follows. In Section \ref{sec:system}, the system model is presented.
In Section \ref{passs1}, the joint power allocation and 3D trajectory design is formulated and solved for  reckless interferer. Section \ref{smarttt} presents the  formulation and the solution of the maximum flow problem for the case of smart interferers. Simulation results are presented in Section \ref{simres} and finally Section \ref{sec:conclusion} concludes the paper.
\section{System Model}\label{sec:system}

In this section, we will describe the communications scenario and the channel models for the air-to-air (A2A) and air-to-ground (A2G) links.

\subsection{Communications Scenario}

\begin{figure}[!t]
	\includegraphics[width=0.45\textwidth]{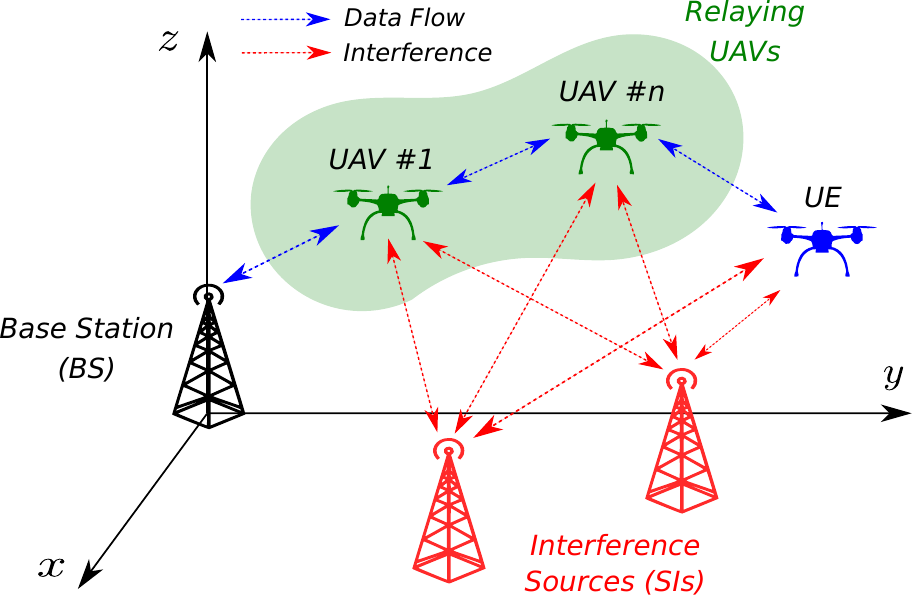}
	\centering
	\caption{System model for the communications scenario where the desired UE is also a flying UAV. The interferers are assumed as reckless fixed interferers.}
	\label{fig:system}
\end{figure}

We consider a scenario where a terrestrial BS and a UE aim to engage in communication. The UE is either  on the ground (e.g., a moving vehicle, pedestrian) or  in the air (e.g., a UAV), as shown in Fig.~\ref{fig:system}. The channel condition of the direct link between
the BS and the UE is not satisfactory for acceptable    
communication performance due to obstacles located in
the line of sight (LoS) area or large distance~\cite{yin2017uav}. To improve the data rate, we consider employing multiple UAVs relaying the signal  between the BS and the UE. We also take into account the interference coming from the existing network (e.g., neighboring BSs, small cells, or malicious jammers), and describe the corresponding transmitters as \textit{interferers}. We term the existing network as the primary network and its UEs as primary UEs. We assume that the interferers can be detected together with their transmission parameters using existing sensing methods in the literature (e.g.,~\cite{tavana2017cooperative}). Moreover, we assume that the BS, the UE, the UAVs and the interferers are functioning as both transmitters and receivers, i.e.,  transceivers, and thus can involve in both uplink and downlink of their own  networks. In addition to the communication-related applications, another use case for this scenario is the aerial wireless sensor networks comprised of UAVs equipped with  sensors and radio devices, which fly over an area of interest to sense and collect data. 

We adopt time-division multiple access (TDMA) to schedule the relaying UAVs so that their  transmissions do not collide with each other. Our  goal is to 
\textit{obtain the 3D trajectories of the relaying UAVs along with the power allocation to maximize the data rate  between the BS and the UE, while the interference constraint on the primary network is met}. As we consider a dynamic network, i.e., the nodes can move, designing 3D trajectories is critical since the UAVs should adaptively reconfigure their locations to avoid the interference and transmit their information simultaneously, more than just  seeking the final locations  to stop and transmit their information. Moreover, 3D trajectory design alone cannot guarantee that the interference threshold constraint is met for the primary network. Thus, a joint 3D trajectory design and power allocation is necessary to address such challenges. 

In our setting, $\mathcal{N}$ describes the set of $N$ nodes in our network, which consists of the terrestrial BS (denoted by node $s$), the desired UE (denoted by node $d$), and the relaying UAVs. In addition, $\mathcal{M}$ stands for the set of $M$ separate interferers. The geometric location of any node in the UAV network  is denoted by $\textbf{r}_i \,{=}\, (x_i,y_i,z_i) \,{\in}\, \mathbb{R}^3$ such that  $i\,{\in}\, \mathcal{N}$. For the interferer nodes, we have $\textbf{r}^J_m \,{=}\, (x^J_m,y^J_m,z^J_m) \,{\in}\, \mathbb{R}^3$ such that $ m\,{\in}\, \mathcal{M}$.  


\subsection{A2A and A2G Channel Models}
 In this section, we discuss the A2A and the A2G channel models under consideration. 
In this work, we consider transmission through the line-of-sight (LoS) path only (e.g., \cite{8424236}), and adopt a widely-used path-loss model provided by the International Telecommunication
Union (ITU). The  path-loss for the A2A link is therefore given (in dB scale) as
\begin{equation}
    \textrm{PL}^{\textrm{A2A}}_{\textrm{dB}}(d_{i,j})=\alpha_1 10 \log_{10} d_{i,j}+ \eta_1 ,
\end{equation}
where $\alpha_1$ is the path-loss exponent, $d_{i,j}$ is the Euclidean distance between the UAVs $i$ and $j$, and $\eta_1$ is the path loss associated with the reference LoS distance of $1\,\text{m}$.
Similarly, the A2G channel between any terrestrial node and the aerial node can be described by the path-loss expression given by
\begin{equation}
    \textrm{PL}^{\textrm{A2G}}_{\textrm{dB}}(d_{i,j})=\alpha_2 10 \log_{10} d_{i,j}+ \eta_2 ,
\end{equation}
where $\alpha_2$ is the path-loss exponent, and $\eta_2$ is the path-loss at the reference distance. Note that in the free space, the path-loss exponents would be $\alpha_\ell\,{=}\,2$, and the path-loss at the reference distance would become $\eta_\ell \,{=}\, 10\log_{10} \,(\frac{4\pi f_{\rm c}}{c})^2$, where $f_{\rm c}$ is the carrier frequency, $c\,{=}\,3\times 10^8\,\text{m/s}$ is the speed of  light, and $\ell \,{\in}\, \{1,2\}$.
As in \cite{ahmed2016importance}, we consider a larger path-loss exponent for the A2G channels  compared to the A2A channels, where these parameters are  presented in Section~\ref{simres}. Considering the impact of small-scale fading, the overall complex channel gain for nodes $i$ and $j$ involved in either the A2A or A2G communications becomes
\begin{equation}
h_{i,j}=\begin{cases}
\displaystyle\frac{g_{i,j}}{\sqrt{\textrm{PL}^{\textrm{A2A}}(d_{i,j})}},& \textrm{if the link is A2A}, \\
\displaystyle\frac{g_{i,j}}{\sqrt{\textrm{PL}^{\textrm{A2G}}(d_{i,j})}}, & \textrm{if the link is A2G},
\end{cases}
\end{equation} 
where $g_{i,j}$ is the small-scale fading gain, which is a zero-mean complex Gaussian random variable with unit variance.  As a final remark, we assume perfect channel reciprocity for all the links under consideration so that the link $(i,j)$ from the node $i$ to node $j$ is equivalent to the link $(j,i)$ in the opposite direction.

 \subsection{Graph Representation of the Network}

\label{probfor}


We assume that   the interference coming from the interferers is much stronger than the noise. We therefore take into account the SIR as the performance metric, which is defined  at node $j$ for the transmission from node $i$ as follows: 
\begin{align}\label{eq:sir}
\textrm{SIR}_{i,j} = \frac{P_i \,  |h_{i,j}|^2}{\sum\limits_{m\in\mathcal{M}} {P}_m^\textrm{J} \,|h_{m,j}|^2+\chi\sum\limits_{k \in {\mathcal{Q}}_{i,j}}u(d_{j,k}/r_{\textrm{int}})},
\end{align}
where  $\mathcal{Q}_{i,j} \,{=}\, \mathcal{N} \,{\backslash}\, \{i,j\}$, $P_i$ is the transmit power of UAV $i$ in the  UAV network.
and ${P}_m^\textrm{J}$ is the transmit power of interferer $m$ in the primary network (i.e., $m\,{\in}\,\mathcal{M}$). Let us define $\mathbf{P}=[P_1, ...,P_N]$ and $\mathbf{P}^J=[P_1^J, ...,P_M^J]$ in vector forms as well.
The second term in the denominator of \eqref{eq:sir} is considered to guarantee a safety separation between any of the UAVs and other nodes in the UAV network (i.e., the BS, the desired UE, or the other UAVs) so as to preserve a proper flight performance~\cite{weibel2004safety}. In this representation, $\chi$ stands for the importance of this safety precaution, and $u$ is the \textit{smoothed} step function given by \cite{6807812}:
\begin{align}\label{eq:smooth_step}
u(y) = \zeta \frac{\textrm{exp}(- \kappa y - \log y_0)}{1+\textrm{exp}(- \kappa y - \log y_0)},    
\end{align}
where $y_0$ is an arbitrarily small positive number, and $\zeta$ and $\kappa$ are design parameters. 
We first define a directed flow graph $G \,{=}\,(\mathcal{N},\mathcal{E})$, in which $\mathcal{E}$ denotes the set of available edges in the network and each edge has the capacity $a_{i,j}$.  We assume a line topology for the multi-hop relay network~\cite{8424236}. This is a reasonable assumption as the major information exchange happens between neighboring UAVs, not between the UAVs away by more than a single hop.  We formulate the information exchange in this single-source and single-destination network as a \textit{single-commodity maximum flow problem}, for which the task is to determine the maximum amount of flow, i.e., the maximum average transmission rate, between the BS and  the desired UE.


The average transmission rate  is defined as the arithmetic mean of the data rates  in the forward and backward directions  for each pair of nodes. The generalized adjacency matrix is accordingly defined as $\mathbf{A} \,{=}\,[a_{i,j}]_{\{i,j\}{=}1}^{N}$, where $a_{i,j}$ is the average transmission rate between nodes $i$ and $j$, given by:
\begin{equation}\label{eq:edge}
a_{i,j}=\begin{cases}
\frac{1}{2}{B} \Big({\log_2(1{+}\textrm{SIR}_{i,j})}{+}{\log_2(1{+}\textrm{SIR}_{j,i})} \Big), & i {\ne} j,\\ 
0, & i{=}j,
\end{cases}
\end{equation}
with ${B}$  the transmission bandwidth of the network. Note that $\textrm{SIR}_{i,j}$ is, in general, not equal to $\textrm{SIR}_{j,i}$, in part, due to the unbalanced deployment of interferer. 
 We further define 
the \textit{generalized degree matrix} of the network as $\mathbf{D}=\textrm{diag}\{\beta_1, ..., \beta_N\}$, where $\beta_i= \sum_{\{j|j \ne i\}}a_{i,j}$. Finally, the \textit{Laplacian matrix} of the network graph is  given by $\mathbf{L} \,{=}\, \mathbf{D} \,{-}\, \mathbf{A}$.

In the following two sections, we formulate the 3D trajectory and power allocation problem introduced in Section under two different types of interference source. In particular, we consider \textit{reckless} and \textit{smart} (i.e., intended) interferers in sequence, and propose a solution to the  problem  for each of these scnarios.

\section{Reckless Interferer }\label{passs1}


We first consider the reckless interference for which the primary role of the interferer is not to produce interference in our UAV-assisted network on purpose, but rather to transmit message signal to its desired UEs, as shown in Fig.~\ref{fig:system}. This type of interference is mainly caused by the transmitters within the existing network (e.g., neighbouring BSs, small cells), and hence is \textit{reckless}. This type of interferer might be static or mobile, and its location is known to the UAVs. Note that any mobile interferer of this type is not moving \textit{intelligently} so as to impair the communications in the UAV-assisted network.

In this scenario, since we assume that the UAVs are working in an already existing network, they must satisfy the interference constraint at primary receivers located on the ground. Thus, we need a constraint on the transmitted signal from the UAVs. On other other hand, for a reliable communication of the primary nodes, they need some order of protection as well. We assume that the interferers (which can serve as BSs in primary network)  are serving a set of primary UEs.
We denote the set of all \textit{\color{black}primary UEs} in the primary network by $\mathcal{R}$, where their  locations are given by $\textbf{r}^\mathcal{R}_u \,{=}\, (x_u,y_u,z_u) \,{\in}\, \mathbb{R}^3$ with $u\,{\in}\, \mathcal{R}$. 
Since the interferers in the reckless case can be assumed as co-existing BSs, we should guarantee that the quality of service (QoS) of its users is satisfied. The interferers can either cooperate with the UAV network or not. In the case of cooperation, the primary network can adjust its transmission power to avoid generating excessive interference to the UAV network while satisfying its own user's QoS. In the absence of cooperation, the interferer in the primary network is not capable of adjusting its transmit power.
We assume that the $\textrm{SINR}$ at primary UEs on the ground is affected by the UAVs interference. The SINR at each UE $u \in \mathcal{R}$ from the transmitter $m \in \mathcal{M}$ while UAV $i$ transmits at the same time is given by 
\begin{equation}
    \textrm{SINR}_{m,u}=\frac{P_m^J \,  |h_{m,u}|^2}{ {P}_i \,|h_{i,u}|^2+\sigma^2},
\end{equation}
where $\sigma^2$ is the noise variance and the first term in the denominator is the interference from the UAV transmitting at the corresponding time slot. The associated rate between the primary UE $u$ and interferer $m$ while the UAV $i$ is transmitting at the same time can therefore be obtained by $R_{m,u}(P_i,P_m^J)=B \log_2(1+\textrm{SINR}_{m,u})$. In order to guarantee the quality of service of the primary UEs, their transmission rates should be larger than a predefined threshold $R^{\rm {th}}$. Considering this assumption, if two networks cooperate, we can adjust the transmit power of the primary network as well to avoid producing more interference while the QoS of its network is guaranteed. In real scenarios, this assumption might not  be able to be satisfied properly; however, it can provide some insights on the upper bound of the performance of the UAV network.

\subsection{Optimization Problem Formulation
}

{\color{black} Here, we formulate the optimization problem for the overall network, where the  goal is to maximize the data flow $\textrm{R}_{\textrm{s} \leftrightarrow \textrm{d}}$ between the terrestrial BS and the desired UE with the help of relaying UAVs in the presence of  interferers.} The flow of the network can be assumed as a measure of average uplink/downlink transmission capability of the UAV network.  
For each link $(i,j) \,{\in}\, \mathcal{E}$, let $f_{i,j}$ be the associated flow such that $0 \,{\le}\, f_{i,j} \,{\le}\, a_{i,j}$. The desired optimization problem is therefore given as follows: 
\begin{IEEEeqnarray}{rl}
\max_{\substack{P_i, \textbf{r}_i \forall i \in \mathcal{Q}_{s,d}\\ {P}_m^\textrm{J}, \forall m \in \mathcal{M}}}
&\qquad \textrm{R}_{\textrm{s} \leftrightarrow \textrm{d}} \;\,{=}\, \!\!\!\sum\limits_{j:(s,j) \in \mathcal{E}} f_{s,j}   \label{eqn:joint_opt_1}\\\hspace{-4mm}
\text{s.t.}
& \!\!\!\sum\limits_{i:(i,j) \in \mathcal{E}} f_{i,j} - \!\!\!\sum\limits_{l:(j,l) \in \mathcal{E}} f_{j,l} \,{=}\, 0, \; \forall  j \in \mathcal{Q}_{s,d}, \IEEEyessubnumber\label{eqn:joint_opt_2}\\
& 0 \le f_{i,j} \le a_{i,j}, ~ \forall (i,j) \in \mathcal{E},  \IEEEyessubnumber\label{eqn:joint_opt_3}\\
&   P_i |h_{i,m}|^2 \le \textrm{I}^{\textrm{max}}_{m}, \; \forall i \in \mathcal{N}, m \in \mathcal{M}, \IEEEyessubnumber\label{eqn:joint_opt_4}\\
& P_i \le \textrm{P}_{\textrm{max}}, \; \forall i \in \mathcal{N},\IEEEyessubnumber\label{eqn:joint_opt_5}\\
& R^{{\rm th}} \le R^i_{u,m}(P_i,P_m^J), ~ \forall m \in \mathcal{M}, u \in \mathcal{R} ,i \in \mathcal{N},~~
\IEEEyessubnumber \label{eqn:joint_opt_6}
\end{IEEEeqnarray}
where $P_i$  stands for the power  of the $i$th UAV, $\textrm{P}_{\textrm{max}}$ is the maximum transmit power of each UAV, and $\textrm{I}^{\textrm{max}}_{m}$ represents the predefined interference threshold for the $m$-th interferer.  In addition, \eqref{eqn:joint_opt_2} is due to the assumption of balanced flows for all the nodes except the  source and the destination. Constraint \eqref{eqn:joint_opt_3} ensures that the flow of each link is  less than the maximum capacity. Moreover, \eqref{eqn:joint_opt_4} satisfies the condition that the interference produced by  each UAV at any interferer is always less than a predefined threshold $\textrm{I}^{\textrm{max}}_{m}$ for the $m$-th interferer. This can guarantee the performance of the primary network in uplink scenario. Constraint $\eqref{eqn:joint_opt_5}$ is imposed to limit the transmission power of the UAVs. Constraint \eqref{eqn:joint_opt_6} is a QoS service consideration for the UEs in the primary network which guarantees the QoS of the primary users in downlink. By imposing this condition and adjusting the transmission power of the primary network interferers, the primary network stops generating stronger interference while satisfying the QoS constraints. 

It is worth mentioning that  given the UAVs' transmission powers, there is no guarantee that the interference constraint is met at the  co-existing primary network with the 3D trajectory design solely~\cite{rahmati2019dynamic}. On the other hand, assuming fixed locations for UAVs, solely optimizing the UAV transmission  powers leads to a poor performance at the UAV relay network. Thus, joint power allocation and 3D trajectory design is necessary to obtain the satisfactory performance for both networks, and it is therefore very complicated to obtain the optimal trajectory and UAVs' transmit powers. In the following, we deploy to decompose the overall optimization of~\eqref{eqn:joint_opt_1} into two sub-problems using the alternating-optimization approach~\cite{bezdek2002some}. In the proposed strategy, we first solve the problem of 3D trajectory optimization for a given set of transmit powers (i.e., $P_i, \forall i \in \mathcal{N}$), and then the power allocation problem is solved for the given set of UAV locations computed beforehand. {\color{black}These recursions continue till a satisfactory level of performance is obtained.} 
\subsection{3D Trajectory Optimization}\label{3Dtraj}

 We first attempt to solve the optimization problem in \eqref{eqn:joint_opt_1} to obtain 3D trajectories of the UAVs assuming an initial set of transmit power values is given.  The optimization therefore reduces to  a maximum flow problem with respect to the locations, which is given as
\begin{IEEEeqnarray}{rl}
\max_{\substack{\textbf{r}_i \\ \forall  i \in \mathcal{Q}_{s,d}}}
&\qquad \textrm{R}_{\textrm{s} \leftrightarrow \textrm{d}} \;\,{=}\, \!\!\!\sum\limits_{j:(s,j) \in \mathcal{E}} f_{s,j}   \label{eqn:unin_loc_opt_1}\\
\text{s.t.}
&\qquad \eqref{eqn:joint_opt_2},  \eqref{eqn:joint_opt_3}. \nonumber 
\end{IEEEeqnarray}
Note that the maximum flow problem in \eqref{eqn:unin_loc_opt_1} can be solved for a given UAV location using the well-known max-flow-min-cut theorem~\cite{ford2015flows}. The achievable maximum flow of the network is equal to single flow min-cut of the underlying network given by
\begin{equation}\label{fflow}
     \textrm{R}^{\textrm{max}}_{\textrm{s} \leftrightarrow \textrm{d}}= \min_{\substack{\{S:v_s\in S, v_d \in \bar S\}}} \sum\limits_{i\in S, j \in \bar S} a_{i,j}.
\end{equation}
The maximum flow in \eqref{fflow} can be obtained by the Ford-Fulkerson algorithm \cite{ford2015flows}. 
The  challenging task is to design the trajectories (i.e., moving directions) of  each UAV in the 3D space so as to maximize the information flow between the BS and the desired UE.   

In order to move towards the maximum flow trajectory, we use \textit{Cheeger constant} or \textit{isoperimetric number} of the graph, which provides numerical measure on how well-connected our multi-node primary wireless network is \cite{chung1997spectral}.  Assuming that $\mathcal{L} \,{=}\, \mathbf{D}^{-1/2}\mathbf{L} \mathbf{D}^{-1/2}$ is the normalized Laplacian matrix, the Cheeger constant is given as \cite{chung1997spectral}
\begin{equation}\label{eq:cheeger}
\displaystyle h(\mathcal{L}) = \underset{S}{\text{min}} \frac {\sum_{i \in S, j \in \bar S}a_{i,j}}{\text{min}\{ |S|,( |\bar S|)\}},
\end{equation}
where $S \,{\subset}\, \mathcal{N}$ is a subset of the nodes, {\color{black} $\bar S \,{=}\, \mathcal{N} \,{-}\, S$}, and $|S|$ is the cardinality of set $S$. 


Note that the original definition of the Cheeger constant $h(\mathcal{L})$ considers all the nodes in the network with equal importance. Since the maximum flow of the network for a given source-destination pair depends on the individual link capacities, the weighted version of the Cheeger constant appears as a promising solution to overcome this drawback. In particular, the original Cheeger constant blindly aims at improving the weakest link in the network and may fail to emphasize the desired flow associated with a particular source-destination pair. 
We therefore need to distinguish between the BS and the desired UE from the UAV nodes, for which the weighted Cheeger constant comes as a remedy, and is given as~\cite{6807812}
\begin{equation}\label{eq:weighted_cheeger}
h_{\mathbf{W}}(\mathcal{L})=\underset{S}{\text{min}} \frac {\sum_{i \in S, j \in \bar S}a_{i,j}}{\text{min}\{ |S|_{\mathbf{W}}),(|\bar S|_{\mathbf{W}})\}},
\end{equation}
where $|S|_{\mathbf{W}} \,{=}\, \sum_{i \in S}w_i$ is the weighted cardinality, and $w_i \,{\ge}\, 0$ is the weight of the node $i$ which is adopted to emphasize any bottleneck along the flow from the BS to the desired UE.
The weighted Laplacian matrix is accordingly given  by
\begin{equation}\label{eq:weighted_laplacian}
\mathcal{L}_\mathbf{W}=\mathbf{W}^{-1/2}\mathcal{L} \mathbf{W}^{-1/2},
\end{equation}
where $\mathbf{W} \,{=}\,{\rm diag} \{w_1, ...,w_n\}$. Usually, the Cheeger constant is difficult to compute. To address this issue, algebraic connectivity can be considered as a suitable alternative. The weighted second smallest eigenvalue $\lambda_2({\mathcal{L}_\mathbf{W}})$ can be defined as
\begin{equation}
\lambda_2(\mathcal{L}_\mathbf{W})= \underset{\textbf{v} \ne \textbf{0}, \textbf{v} \perp \mathbf{W}^{1/2}\textbf{1}}{\text{min}} \frac{ \langle \mathcal{L}_\mathbf{W} \textbf{v}, \textbf{v} \rangle}{\langle \textbf{v}, \textbf{v}\rangle}. 
\end{equation}
It is shown in \cite{6807812} that the following weighted Cheeger's inequalities hold
\begin{equation}
\lambda_2(\mathcal{L}_\mathbf{W})/2 \le h_\mathbf{W}(\mathcal{L}_\mathbf{W}) \le \sqrt{2 \delta_{\textrm{max}} \lambda_2(\mathcal{L}_\mathbf{W})/w_{\textrm{min}}} \,,
\end{equation}
where $\delta_{\textrm{max}}$ is the maximum node degree, 
and $w_{\textrm{min}}= \min_i w_i$. As can be seen, when $\lambda_2(\mathcal{L}_\mathbf{W})$ gets larger values, the lower bound of the weighted Cheeger constant increases, which improves the connectivity of the overall network, and suppresses the formation of bottleneck. The UAVs can therefore adjust their geometric locations in order to maximize $\lambda_2({\mathcal{L}_\mathbf{W}})$, and hence $h_{\mathbf{W}}(\mathcal{L})$. 

As a result, each UAV should move along the \textit{spatial} gradient of the weighted algebraic connectivity $\lambda_2({\mathcal{L}_\mathbf{W}})$ to maximize it. Given the instantaneous location of the $i$th UAV, its spatial gradient along $x$-axis is given as follows:
\begin{align}
\frac{\partial \lambda_2({\mathcal{L}_\mathbf{W}}) }{\partial x_i} & =    {\mathbf{x}^f}^T \frac{\partial ({{\mathcal{L}_\mathbf{W}}}) }{\partial x_i} {\mathbf{x}^f} = \sum_{p=1}^N \sum_{q=1}^N\frac{v_p^f}{\sqrt{w_p}}\frac{v_q^f}{\sqrt{w_q}}\left[\frac{\partial \mathcal{L} }{\partial x_i}\right]_{p,q}\\
&= \sum_{\{p,q:p \sim q \}}  \left[\frac{v_p^f}{\sqrt{w_p}}-\frac{v_q^f}{\sqrt{w_q}}\right]^2\frac{\partial a_{p,q}}{\partial x_i}, & 
\label{eq:spatial_gradient_unintended_2}
\end{align}
where  ${v}^f_k$ is the $k$th entry of $\mathbf{v}^f$ with $k\,{\in}\,\{p,q\}$, $\mathbf{v}^f$ is the Fiedler vector which is the eigenvector corresponding to the second smallest eigenvalue {\color{black}$\lambda_2({\mathcal{L}_\mathbf{W}})$}, and $p\,{\sim}\,q$ means that
the nodes $p$ and $q$ are connected. In  \eqref{eq:spatial_gradient_unintended_2},  $\frac{\partial a_{p,q}}{\partial x_i}$ can be computed  as
\begin{align}\label{eq:partial_derivative}
\frac{\partial a_{p,q}}{\partial x_i} &= \textrm{B}
\left[ \frac{1}{1{+}\textrm{SIR}_{p,q}}\frac{\partial \textrm{SIR}_{p,q}}{\partial x_i}  + \frac{1}{1{+}\textrm{SIR}_{q,p}}\frac{\partial \textrm{SIR}_{q,p}}{\partial x_i} \right],
\end{align}
which is $0$ for $p\,{=}\,q${\color{black}, or $i \,{\notin}\,\{p,q\}$}. The partial derivative of SIR with respect to $x_i$ in \eqref{eq:partial_derivative} can be computed using \eqref{eq:sir} together with the geometrical relations between $x_i$ and the complex channel gain $h_{i,j}$ presented in Section~\ref{sec:system}.
Assuming $p\,{=}\,i$ and $p\neq q$, we have 
\begin{align}
 \frac{\partial \textrm{SIR}_{i,q}}{\partial x_i} = \frac{10^{-\eta_l/10}|g_{i,q}|^2\alpha_ld_{i,q}^{-\alpha_l-2}(x_q-x_i)}{{\sum\limits_{m\in\mathcal{M}} {P}_m^\textrm{J} \,|h_{q,m}|^2+\chi\sum\limits_{k \in {\mathcal{Q}}_{i,j}}u(d_{q,k}/r_{\textrm{int}})}},
\end{align}
and 
\begin{align}\nonumber
 \frac{\partial \textrm{SIR}_{q,i}}{\partial x_i} =&  \frac{10^{-\eta_l/10}|g_{i,q}|^2\alpha_ld_{i,q}^{-\alpha_l-2}(x_q-x_i)}{\Big({\sum\limits_{m\in\mathcal{M}} {P}_m^\textrm{J} \,|h_{i,m}|^2+\chi\sum\limits_{k \in {\mathcal{Q}}_{i,j}}u(d_{i,k}/r_{\textrm{int}})}\Big)} \nonumber \\
+ & \frac{ d^{-\alpha_l}_{i,m} 10^{-\eta_l/10}\sum\limits_{m\in\mathcal{M}}|g_{i,m}|^2\alpha_ld_{i,m}^{-\alpha_l-2}(x_i-x_m)}{\Big({\sum\limits_{m\in\mathcal{M}} {P}_m^\textrm{J} \,|h_{i,m}|^2+\chi\sum\limits_{k \in {\mathcal{Q}}_{i,j}}u(d_{i,k}/r_{\textrm{int}})}\Big)^2}  \nonumber \\
+ & \frac{ d^{-\alpha_l}_{i,m} 10^{-\eta_l/10}\sum\limits_{k \in {\mathcal{Q}}_{i,j}}u^{'}(d_{i,k}/r_{\textrm{int}})\frac{x_k-x_i}{r_\textrm{int}d_{i,k}}}{\Big({\sum\limits_{m\in\mathcal{M}} {P}_m^\textrm{J} \,|h_{i,m}|^2+\chi\sum\limits_{k \in {\mathcal{Q}}_{i,j}}u(d_{i,k}/r_{\textrm{int}})}\Big)^2},
\end{align}
where $u^{'}(x)\,{=}\,{\rm{d}} u/{\rm{ d}} x$. If 
$i \,{\notin}\,\{p,q\}$ and $p \,{\ne}\, q$, one can obtain
\begin{align}\label{p1q}
 \frac{\partial \textrm{SIR}_{p,q}}{\partial x_i} =&  \frac{u^{'}(d_{q,i}/r_{\textrm{int}})}{\Big({\sum\limits_{m\in\mathcal{M}} {P}_m^\textrm{J} \,|h_{q,m}|^2+\chi\sum\limits_{k \in {\mathcal{Q}}_{i,j}}u(d_{q,k}/r_{\textrm{int}})}\Big)^2}\frac{x_i-x_q}{r_\textrm{int}d_{q,i}},
\end{align}
and ${\partial \textrm{SIR}_{q,p}}/{\partial x_i}$ can be obtained similar to \eqref{p1q} by changing the subscripts $p$ and $q$.
The update in the location of the $i$th UAV along the $x-$axis is then given as
\begin{align}\label{eq:loc_update_x}
    x_i(t+1)=x_i(t)+{\rm d}t \frac{\partial \lambda_2({\mathcal{L}_\mathbf{W}}) }{\partial x_i(t)},
\end{align}
where $t$ stands for the discrete time, or, equivalently, the iteration number. A similar procedure can be pursued to find the spatial gradients of $\lambda_2({\mathcal{L}_\mathbf{W}})$ along the  $y-$ and $z-$axis, and update the  coordinates. 




\subsection{Power Allocation Optimization}\label{powallo}

We now focus on the power allocation problem, and solve the optimization problem in \eqref{eqn:joint_opt_1} to find the optimal power allocation for a given set of the UAV locations. In this case, we consider the reckless   interferer  generating the unwanted interference. Here, we assume that the primary network and the  UAV network can cooperate so as to mitigate the mutual interference. The UAVs try to mitigate the mutual unwanted interference between the primary co-existing network and the UAV network in order to improve their information flow between the source and the destination. However, 3D trajectory design solely can not guarantee that the interference threshold is met in the primary network. Thus, the power allocation should be done in order to guarantee the interference threshold constraint. The corresponding optimization becomes 
\begin{IEEEeqnarray}{rl}
\max_{\substack{P_i  \forall i \in \mathcal{Q}_{s,d}\\ {P}_m^\textrm{J}, \forall m \in \mathcal{M}}}
&\qquad \textrm{R}_{\textrm{s} \leftrightarrow \textrm{d}} \;\,{=}\, \!\!\!\sum\limits_{j:(s,j) \in \mathcal{E}} f_{s,j}   \label{eqn:unin_pow_opt_1}\\
\text{s.t.}
&\qquad \!\eqref{eqn:joint_opt_2},  \eqref{eqn:joint_opt_3}, \eqref{eqn:joint_opt_4}, \eqref{eqn:joint_opt_5},
\eqref{eqn:joint_opt_6}.\nonumber 
\end{IEEEeqnarray}
In this case, the maximum information exchange of the network is determined by the link with the minimum instantaneous rate.
Hence, as we try to maximize the flow of the network, we equivalently maximize the information exchange of the hop with the minimum instantaneous rate. 
More specifically, the power allocation problem can be equivalently given by 
\begin{IEEEeqnarray}{rl}\label{eqqe}
\max_{\substack{P_i  \forall i \in \mathcal{Q}_{s,d}\\ {P}_m^\textrm{J}, \forall m \in \mathcal{M}}} \;  \min_{\substack{ \forall \, j \in \mathcal{Q}_{s,d}}}
&\qquad a_{i,j}  \label{eqn:unin_pow_opt_2}\\
\text{s.t.}
&\qquad \eqref{eqn:joint_opt_4}, \eqref{eqn:joint_opt_5},
\eqref{eqn:joint_opt_6}. \nonumber 
\end{IEEEeqnarray}
In order to have a more tractable problem, \eqref{eqn:unin_pow_opt_2} can be reformulated as follows:
\begin{IEEEeqnarray}{rl}\label{optww}
\max_{\substack{P_i  \forall i \in \mathcal{Q}_{s,d}\\ {P}_m^\textrm{J}, \forall m \in \mathcal{M}}} & \qquad \eta \label{eqn:unin_pow_opt_3}\\
\text{s.t.}
&\qquad 0 \le \eta \le a_{i,j}, \,\forall j \in \mathcal{Q}_{s,d}, \IEEEyessubnumber \label{eqn:unin_pow_opt_31} \\
&\qquad \eqref{eqn:joint_opt_4}, \eqref{eqn:joint_opt_5},
\eqref{eqn:joint_opt_6}, \nonumber 
\end{IEEEeqnarray}
where $\eta$ is an auxiliary variable employed to facilitate the optimization. In this case, the objective is an affine function. However, the constraint on  $a_{i,j}$ is not convex.  $a_{i,j},$ $\forall \, i,j \in \mathcal{Q}_{s,d}$, can  
be recast as the  difference of two concave functions, given by $a_{i,j}(P_i, P_j,\mathbf{P}^J)= v(P_i, P_j,\mathbf{P}^J) -r(\mathbf{P}^J)$,  where
\begin{align}
  &v(P_i, P_j,\mathbf{P}^J) = \nonumber \\ &\frac{1}{2} B\log_2\left({P_i \,  |h_{i,j}|^2}+ \hspace{-1.5mm}{\sum\limits_{m\in\mathcal{M}} {P}_m^\textrm{J} \,|h_{m,j}|^2 +\chi \hspace{-3mm}\sum\limits_{k \in {\mathcal{Q}}_{i,j}}u(d_{j,k}/r_{\textrm{int}})}\right)+ \nonumber\\ 
  & \frac{1}{2} B\log_2\left({P_j \,  |h_{j,i}|^2}+\hspace{-1.5mm}{\sum\limits_{m\in\mathcal{M}}\hspace{-1.5mm} {P}_m^\textrm{J} \,|h_{m,i}|^2 +\chi \hspace{-3mm}\sum\limits_{k \in {\mathcal{Q}}_{i,j}}u(d_{i,k}/r_{\textrm{int}})}\right), 
\end{align}
and  
\begin{align}
 r(\mathbf{P}^J)= &\frac{1}{2} B \log_2\Big({\sum\limits_{m\in\mathcal{M}} {P}_m^\textrm{J} \,|h_{m,j}|^2  +\chi\sum\limits_{k \in {\mathcal{Q}}_{i,j}}u(d_{j,k}/r_{\textrm{int}})}\Big)\nonumber\\ 
 + &  \frac{1}{2} B \log_2\Big({\sum\limits_{m\in\mathcal{M}} {P}_m^\textrm{J} \,|h_{m,i}|^2  +\chi\sum\limits_{k \in {\mathcal{Q}}_{i,j}}u(d_{i,k}/r_{\textrm{int}})}\Big).
\end{align}
\begin{algorithm}[t]
\caption{DC-based SCA Algorithm for Power Allocation}
\begin{algorithmic}[1]
\State \textbf{Initialization:} Iteration number $t:=1$, locations $\textbf{r}_i$  $, \forall i \in \mathcal{N}$, feasible initial value for  $P_i[0],~\forall i \in \mathcal{N}$ and $\mathbf{P}^J{[0]}$.
\State Calculate the value of $  \tilde a_{i,j}(P_i[0], P_j[0],\mathbf{P}^J[0])$, $\forall i \in \mathcal{N}$ using~\eqref{f-approx}.
\While{$|\eta[t] \,{-}\, \eta[t-1]| \,{>}\, \varepsilon$ }

  \State Compute the optimal power allocation  $P_i[t],~\forall i \in \mathcal{N}$ and  $\mathbf{P}^J{[t]}$ in \eqref{eeeeq} using CVX \cite{grant2014cvx}.
  
  \State Compute  $ \tilde  a_{i,j}(P_i[t], P_j[t],\mathbf{P}^J[t])$, $\forall i \in \mathcal{N}$ using~\eqref{f-approx}. 

\State $t \gets t + 1$
\EndWhile
\end{algorithmic}\label{alg:alg1}
\end{algorithm}
 In general, the difference of two concave functions is not a concave one~\cite{chiang2005power}. In order to convexify this function at iteration $t$, we deploy first-order  Taylor expansion to approximate $r(\mathbf{P}^J)$  around a given point from the previous iteration $\mathbf{P}^J[t-1]$ as:
\begin{equation}\label{eqapp}
    \tilde r(\mathbf{P}^J)\approx r(\mathbf{P}^J[t-1])+(\nabla r(\mathbf{P}^J[t-1]))^T(\mathbf{P}^J-\mathbf{P}^J[t-1]).
\end{equation}
Thus, we have the approximated version of $a_{i,j}(P_i, P_j,\mathbf{P}^J)$ as 
\begin{equation}\label{f-approx}
    \tilde a_{i,j}(P_i, P_j,\mathbf{P}^J)=v(P_i, P_j,\mathbf{P}^J)-\tilde r(\mathbf{P}^J).
\end{equation}
Using this approximation,  one can see that $\tilde a_{i,j}(P_i, P_j,\mathbf{P}^J)$ is a concave function. A similar approximation can be adopted for the constraint $\eqref{eqn:joint_opt_6}$ to make it convex as a new constraint~($\hat{\textrm{8e}}$).
To do this, we can re-write the constraint  $\eqref{eqn:joint_opt_6}$ as:
Thus, the optimization problem in \eqref{optww} can be recast as:
\begin{IEEEeqnarray}{rl}\label{eeeeq}
\max_{\substack{P_i  \forall i \in \mathcal{Q}_{s,d}\\ {P}_m^\textrm{J}, \forall m \in \mathcal{M}}} & \qquad \eta \label{eqn:unin_pow_opt_311}\\
\text{s.t.}
&\qquad 0 \le \eta \le \tilde a_{i,j}, \,\forall j \in \mathcal{Q}_{s,d}, \IEEEyessubnumber \label{eqn:unin_pow_opt_312} \\
&\qquad \eqref{eqn:joint_opt_4}, \eqref{eqn:joint_opt_5},
(\hat{\textrm{8e}}).\nonumber 
\end{IEEEeqnarray}
After substituting the approximated versions of the constraints, in each iteration $l$, the above optimization problem is now convex, which can be solved efficiently using the interior point method~\cite{boyd2004convex}. This procedure,  called successive convex approximation (SCA)~\cite{nasir2016joint}, is described in Algorithm~\ref{alg:alg1}.

\begin{proposition}
Algorithm 1 generates a sequence of improved feasible points that converge to a point $(\mathbf{P}^*, \mathbf{P}^{J*})$ satisfying the KKT conditions of the problem \eqref{eqn:unin_pow_opt_3}.
\end{proposition}
\subsubsection*{Proof} The proof is omitted due to brevity. Similar proof can be found in \cite{khamidehi2016joint}.


If the  primary network does not cooperate with UAV network to adjust its transmission power so as to help  improve the UAV transmission flow, the optimization can be done with respect to the UAVs' transmit powers while the  transmit powers of the existing network  are fixed.


\subsection{The Joint Alternating-Optimization Algorithm}
Given the transmit powers of the UAVs, the trajectory design problem can be solved based on the  Cheeger constant in Section~\ref{3Dtraj}. Given the locations of the UAVs, the power allocation can be obtained using the  SCA method as in Algorithm 1 discussed in Section~\ref{powallo}. At each step of 3D trajectory design, we need to make sure that the interference threshold constraint is met in the primary network. Thus, at each iteration of 3D trajectory design, we compute new set of  transmit powers for UAVs. The overall algorithm considering both the 3D trajectory and power allocation optimization is summarized in Algorithm~\ref{alg:alg}. 
\begin{algorithm}
\caption{Proposed Alternating-Optimization Algorithm}
\begin{algorithmic}[t]
\State \textbf{Initialization:} Locations $\textbf{r}_i$ and power allocation $P_i$ for $i$th UAV for $\forall i \,{\in}\,\mathcal{N}\,{\cup}\,\mathcal{M}$, error tolerance $\varepsilon$
\State $t \gets 1$, $\textrm{R}_{\textrm{s} \leftrightarrow \textrm{d}}({-}1) \gets {-}\infty$, $\textrm{R}_{\textrm{s} \leftrightarrow \textrm{d}}(0) \gets 0$
\While{ $|\textrm{R}_{\textrm{s} \leftrightarrow \textrm{d}}(t{-}1) \,{-}\, \textrm{R}_{\textrm{s} \leftrightarrow \textrm{d}}(t{-}2)| \,{>}\, \varepsilon$ }
\State Compute $\frac{\partial \lambda_2({\mathcal{L}_\mathbf{W}})}{\partial x_i(t)}$, $\frac{\partial \lambda_2({\mathcal{L}_\mathbf{W}})}{\partial y_i(t)}$, $\frac{\partial \lambda_2({\mathcal{L}_\mathbf{W}})}{\partial z_i(t)}$ by \eqref{eq:spatial_gradient_unintended_1}-\eqref{eq:partial_derivative}
\State Update $\textbf{r}_i(t)$ in $\mathbb{R}^3$ as in \eqref{eq:loc_update_x} 
\State Compute optimal $\mathbf{P}$ and $\mathbf{P}^J$  using SCA algorithm given in Algorithm 1. 
\State Compute $\textrm{R}_{\textrm{s} \leftrightarrow \textrm{d}}(t)$ by  Ford-Fulkerson algorithm~\cite{ford2015flows}
\State $t \gets t + 1$
\EndWhile
\end{algorithmic}\label{alg:alg}
\end{algorithm}
\section{ Smart Interferer }\label{smarttt}
\begin{figure}[!t]
	\includegraphics[width=0.45\textwidth]{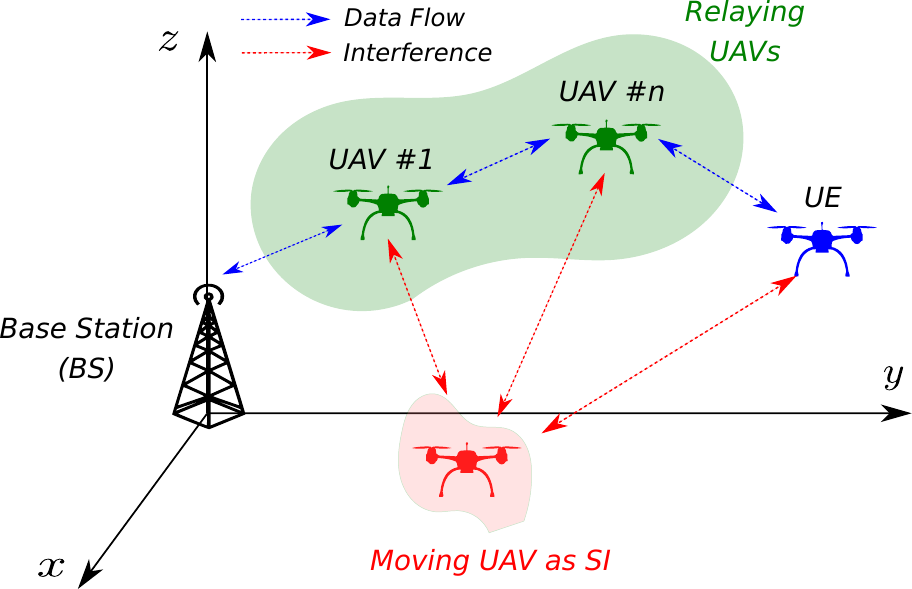}
	\centering
	\caption{System model for the communications scenario where the interferer is a smart moving UAV.}
	\label{fig:system2}
\end{figure}

In this scenario, we assume that interferers are smart, as shown in Fig.~\ref{fig:system2}. Smart interferers can move in order to decrease the flow of the UAV network. The moving smart interferers can be assumed as other UAVs trying to interrupt or at least degrade the  quality of the  communication link for the legitimate UAVs. In this case, the UAVs act selfishly to improve their own transmission quality which means the interference constraint is dropped.  
We consider the problem from both the UAV network and smart interferer's perspective. The UAVs in the UAV network try to reconfigure their 3D locations to evade the interference caused by the smart interferers, while the smart interferers' goal is to chase the UAVs to decrease the path-loss effect and hence increase the intended interference to the UAV network. In this case, the UAVs can transmit with maximum power. Hence,  the optimization problem is only over trajectory design as
\begin{IEEEeqnarray}{rl}
\max_{\substack{ \textbf{r}_i \forall i \in \mathcal{Q}_{s,d}}}
&\qquad \textrm{R}_{\textrm{s} \leftrightarrow \textrm{d}} \;\,{=}\, \!\!\!\sum\limits_{j:(s,j) \in \mathcal{E}} f_{s,j}   \label{eqn:joint_opt_122}\\
\text{s.t.}
&\qquad \!\!\!\sum\limits_{i:(i,j) \in \mathcal{E}} f_{i,j} - \!\!\!\sum\limits_{l:(j,l) \in \mathcal{E}} f_{j,l} \,{=}\, 0, \; \forall  j \in \mathcal{Q}_{s,d}, \IEEEyessubnumber\label{eqn:joint_opt_222}\\
&\qquad 0 \le f_{i,j} \le a_{i,j}, ~ \forall (i,j) \in \mathcal{E}.
\IEEEyessubnumber \label{eqn:joint_opt_6222}
\end{IEEEeqnarray}
The above optimization problem is over multiple UAVs. In the following, we address the problem from both the perspective of UAV network and smart jammers.
\subsection{From UAV Network Perspective}
In the UAV network, the UAVs try to evade the smart interferers' interference. For the power allocation, since there are no constraints on the interference of the UAVs to smart jammers, the best strategy is to transmit with full power.  
In  the 3D trajectory design, the UAVs deploy the Cheeger constant metric  and move toward the spatial gradients of the algebraic connectivity so as to maximize the flow of the network as in~\eqref{eq:spatial_gradient_unintended_2}. 

\subsection{From the Smart Interferers' Perspective}
Here, we  consider the problem from the perspective of  moving smart interferers which can be assumed as other UAVs aiming at  decreasing the flow of the network. To do this, we assume that the smart UAVs  transmit with their full power and move towards the opposite direction of spatial gradient of the algebraic connectivity of the UAV network weighted Laplacian matrix. Thus, the moving direction for the moving smart interferers can be given by: 
\begin{align}
\frac{\partial \lambda_2({\mathcal{L}_\mathbf{W}}) }{\partial x^J_m} & =  -  {\mathbf{x}^f}^T \frac{\partial ({{\mathcal{L}_\mathbf{W}}}) }{\partial x^J_m} {\mathbf{x}^f} \label{eq:spatial_gradient_unintended_1}=
-\sum_{p=1}^N \sum_{q=1}^N\frac{v_p^f}{\sqrt{w_p}}\frac{v_q^f}{\sqrt{w_q}}\left[\frac{\partial \mathcal{L} }{\partial x_m^J}\right]_{p,q}\\
&=- \sum_{\{p,q:p \sim q \}}  \left[\frac{v_p^f}{\sqrt{w_p}}-\frac{v_q^f}{\sqrt{w_q}}\right]^2\frac{\partial a_{p,q}}{\partial x^J_m}, ~\forall~ m \in \mathcal{M}. & 
\label{eq:spatial_gradient_unintended22}
\end{align}
By moving along the opposite direction of spatial gradient of the  second smallest eigenvalue of the weighted Laplacian matrix of the UAV network, the moving smart interferers can  decrease the maximum flow between the BS and the UE. The partial derivative with respect $x_m^J, ~\forall~ m \in \mathcal{M}$ can be obtained similar to \eqref{eq:partial_derivative} as follows: 
\begin{align}\label{pdd44q}
 \frac{\partial \textrm{SIR}_{i,j}}{\partial x^J_m} =&  \frac{10^{-\eta_l/10}|g_{j,m}|^2\alpha_l d_{j,m}^{-\alpha_l-2}(x_j-x^J_m)}{\Big({\sum\limits_{m\in\mathcal{M}} {P}_m^\textrm{J} \,|h_{j,m}|^2+\chi\sum\limits_{k \in {\mathcal{Q}}_{i,j}}u(d_{j,k}/r_{\textrm{int}})}\Big)^2}P_i \,  |h_{i,j}|^2.
\end{align}
For $y$-axis and $z$-axis, similar equations can be obtained. We consider a parameter $\tau$ so as to adjust the level of smartness of the smart UAV. That means the smart UAV interferer can move every $\tau$ iterations. Thus, by decreasing the value of $\tau$ the moving interferer will be smarter as it can chase the relay UAVs faster. It should be noted that we consider the static interferer as naive interferer in this scenario.

\begin{table}\caption {Simulation Parameters} \label{tab:simulation} 
\renewcommand{\arraystretch}{1.0}
\centering
\begin{tabular}{ lc }
\hline
Parameter & Value \\
\hline
\hline
Path-loss exponents $(\alpha_1, \alpha_2)$ & $2.05, 2.32$  \\
Maximum transmit power of the UAVs $(\textrm{P}_{\textrm{max}})$ & $20\,\text{dBm}$\\
Transmit power of the interferers  $(\textrm{P}_m^\textrm{J}$, $\forall m\,{\in}\,\mathcal{M})$ & $30\,\text{dBm}$ \\
Bandwidth $(\textrm{B})$ & $10\,\text{KHz}$ \\ 
Interference threshold $(\textrm{I}^{\textrm{th}}_m$, $\forall m\,{\in}\,\mathcal{M})$ & $[{-}50,{-}10]\,\text{dBm}$ \\ 
Interference radius $(r_{\textrm{int}})$ & $5\,\text{m}$ \\ 
Carrier frequency $(f_{\rm c})$ & $2~ \text{GHz}$ \\ 
Smoothed step-function parameters $(\zeta,\kappa,{\color{black}y_0})$ & $1,10, 10^{-3}$ \\
Safety precaution priority $(\chi)$ & $1$ \\
\hline
\end{tabular}
\end {table}

\section{Simulation Results}\label{simres}
\begin{figure}[!t]
	\centering
	\includegraphics[width=0.52\textwidth]{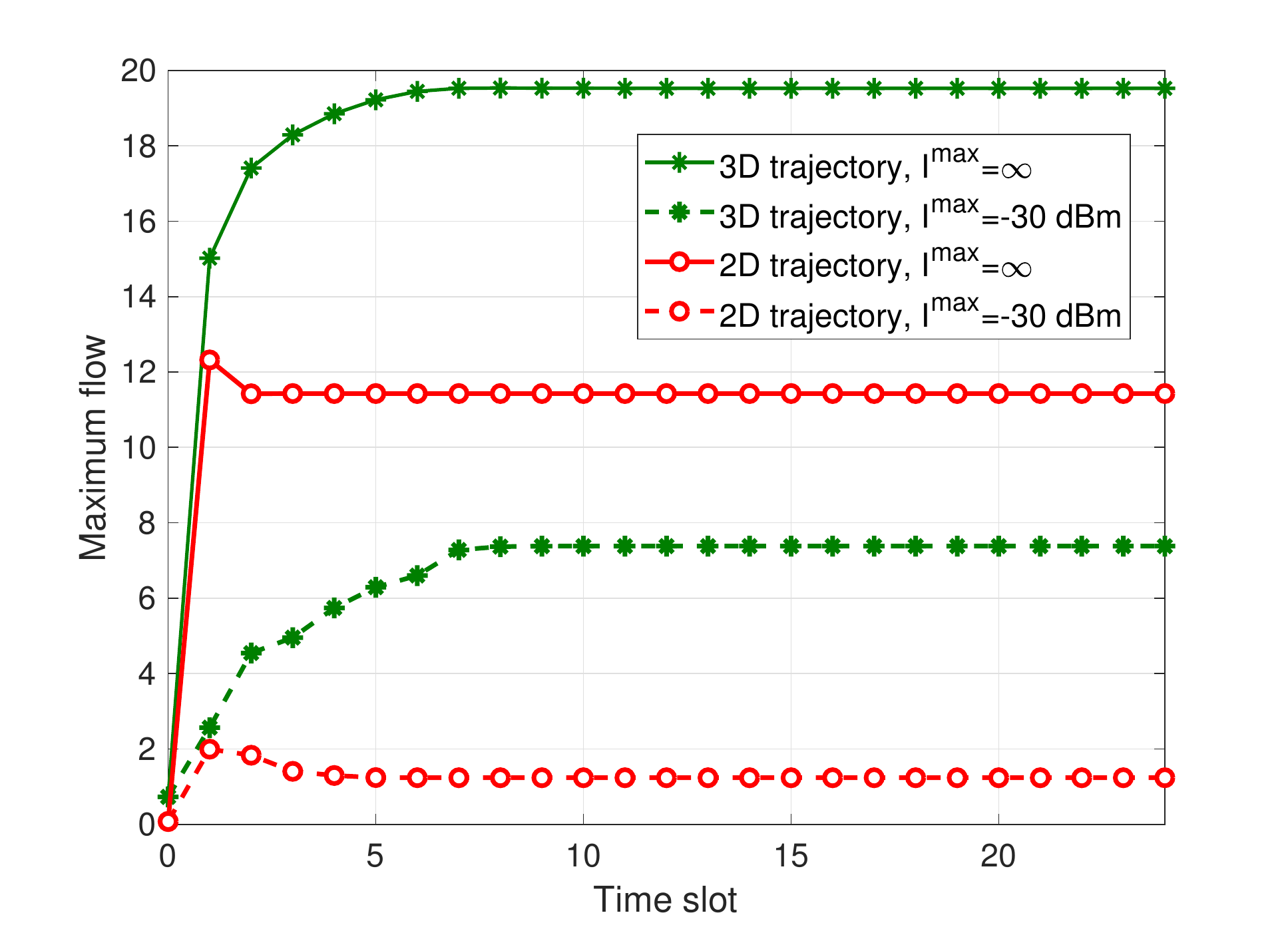}
	\caption{Convergence of the  proposed alternating optimization problem.}
	\label{fig:flow_vs_iteration}
\end{figure}

In this section, we present numerical results based on extensive simulations, to evaluate the performance of the proposed joint 3D trajectory and power allocation optimization. In our simulation environment, the BS and the UE are assumed to be located at $(0,0,h^\textrm{BS})$ and $(200\,\text{m},0,h^\textrm{UE})$, respectively, in $\mathbb{R}^3$ with $h^\textrm{BS} \,{=}\,15\,\text{m}$. Moreover, the reckless interferers are located randomly in $xy-$plane with fixed altitude of $h^\textrm{SI} \,{=}\,20\,\text{m}$. The list of simulation parameters are given in Table~\ref{tab:simulation}.

\subsection{Reckless Interferer}
Here, we assume that the interferer acts as an reckless interferer which can be interpreted as a  transceiver in the co-existing network.

\begin{figure}[!t]
	\centering
	\includegraphics[width=0.48\textwidth]{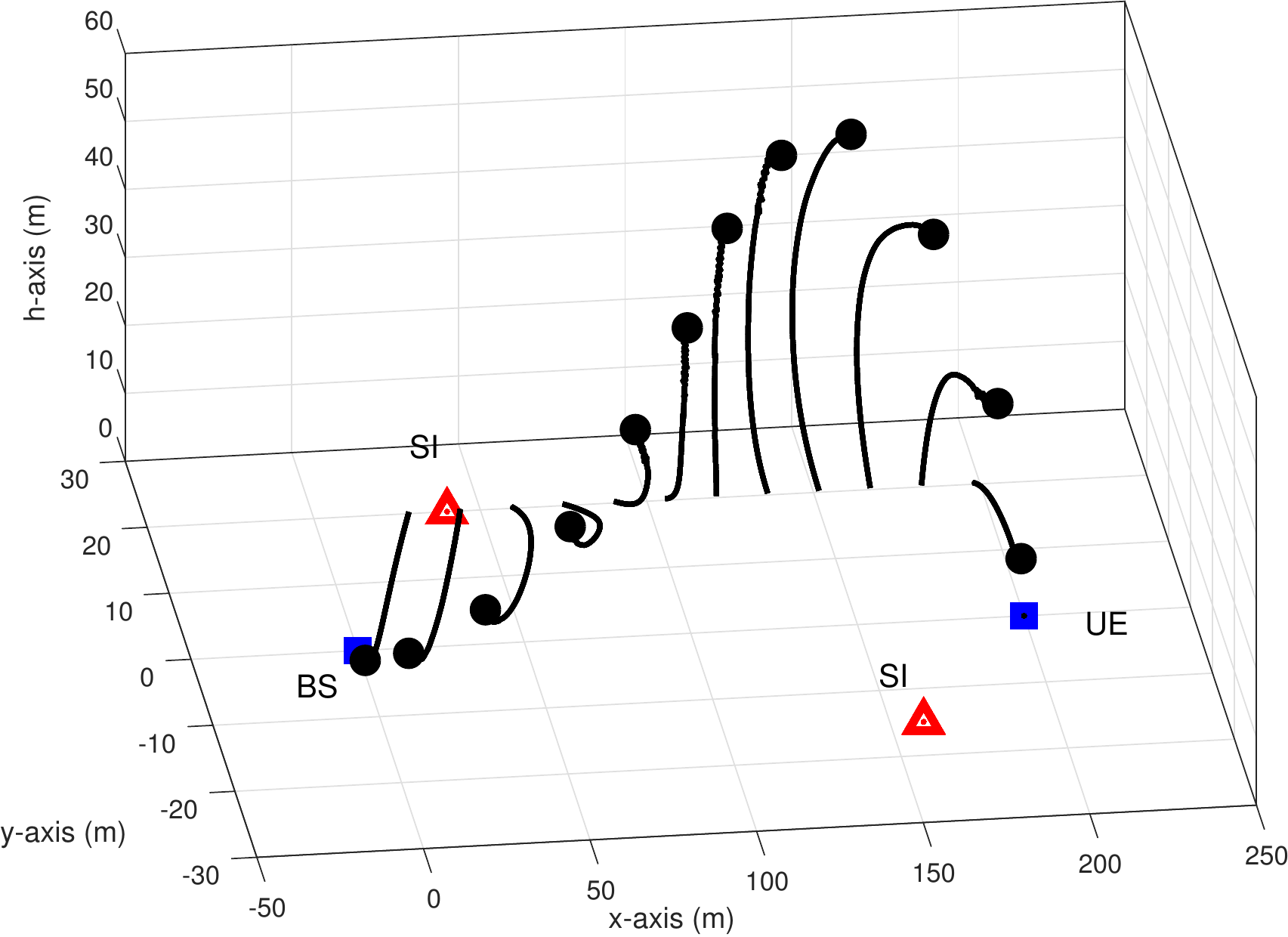}
	\caption{3D trajectories of the UAVs in the presence of two reckless interferers (3D view).}
	\label{fig:3d}
\end{figure}
\begin{figure}[!t]
	\centering
	\includegraphics[width=0.45\textwidth]{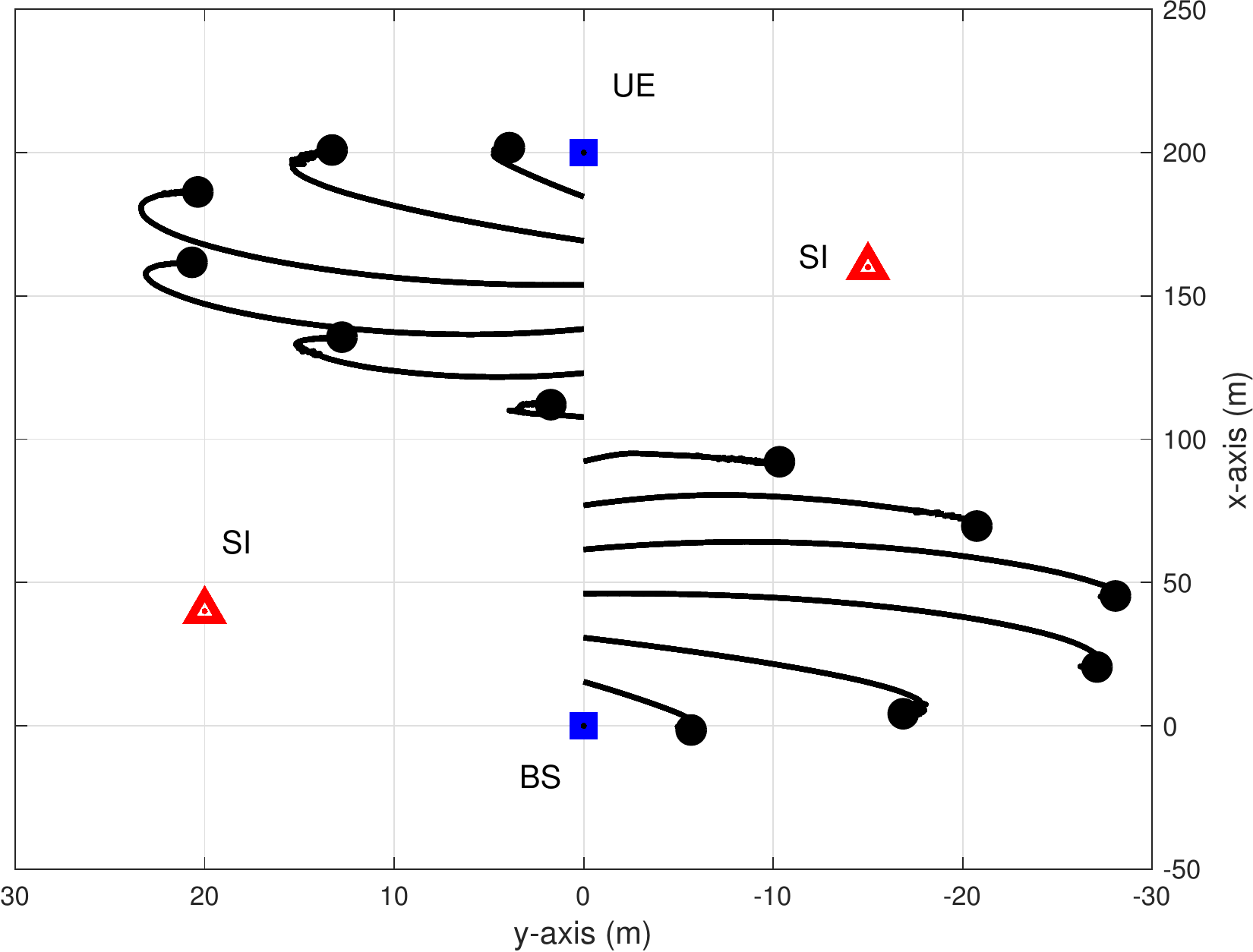}
	\caption{3D trajectories of the UAVs in the presence of two reckless interferers (top view).}
	\label{fig:top}
\end{figure}

\subsubsection{Convergence of the Proposed Algorithm}
In Fig. \ref{fig:flow_vs_iteration}, the maximum flow of the network is depicted versus iterations. Each iteration is composed of solving one power allocation optimization problem and one trajectory design. We plotted the results for different value of $I^{\max}$, which determines the maximum value of tolerable interference form UAV network on the primary network. Apparently, the larger the $I^{\max}$, the larger data flow can be sent from the BS to UE as the UAVs can transmit with more power.
We show  the  maximum flow of the network for 2D and 3D trajectory design approaches. For the $xy-$plane 2D trajectory design, we assume that each UAV can move in $xy-$plane and it can not move in $z$ direction (its height is assumed fixed).  It can be seen that both algorithms  converge in finite number of iterations, while the 3D trajectory design needs more time for convergence. However, 3D trajectory design  significantly outperforms the 2D  trajectory design and can double the transmission flow of the network. This performance improvement, which is one strong benefit of 3D trajectory design, is due to the fact that 3D space has more degrees of freedom  as compared to 2D.

\begin{figure}[!t]
	\centering
	\includegraphics[width=0.43\textwidth]{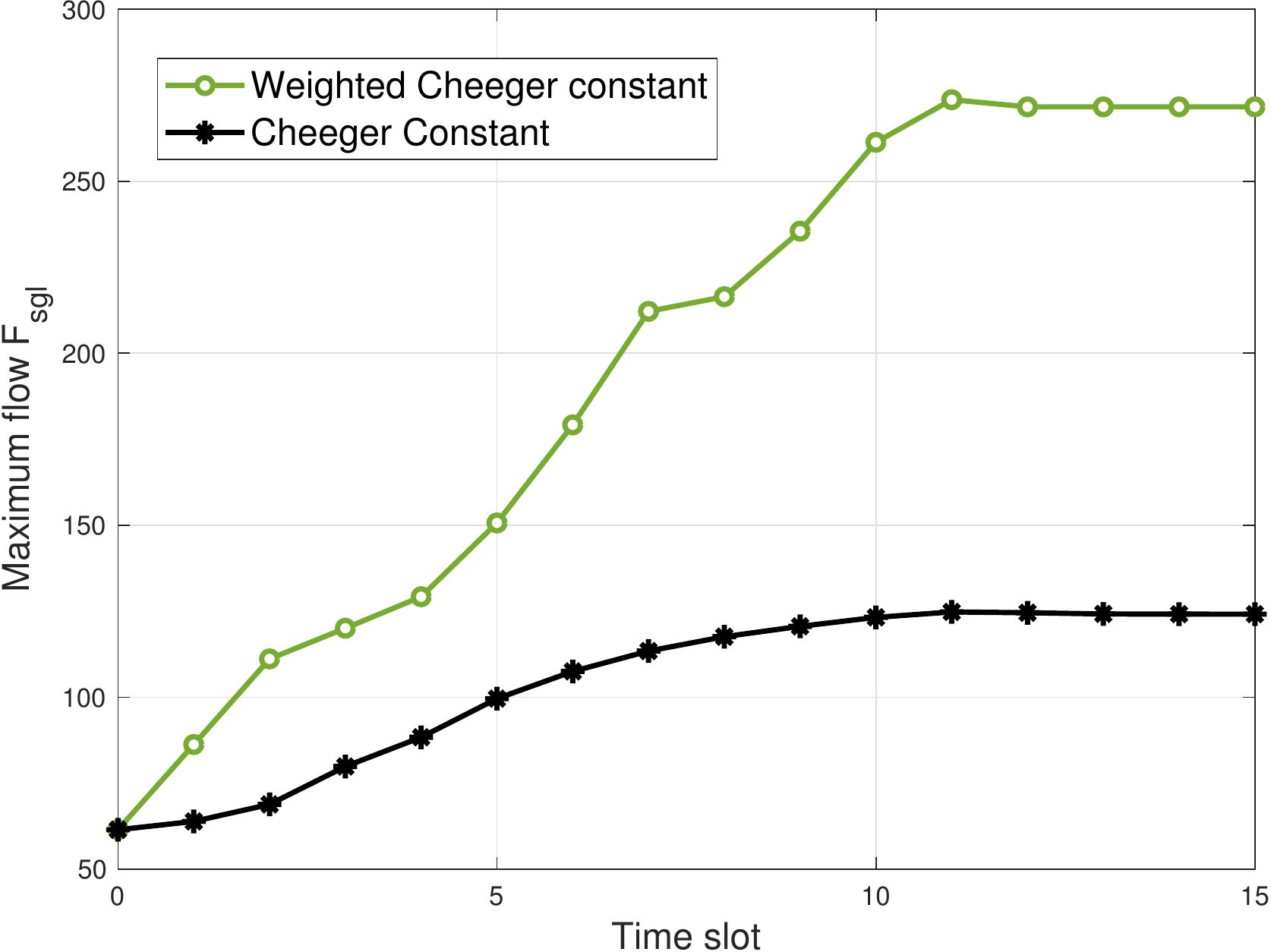}
	\caption{Maximum flow of the network for weighted and unweighted Cheeger constant.}
	\label{cheegflow}
\end{figure}

\begin{figure}[!t]
	\centering
	\includegraphics[width=0.48\textwidth]{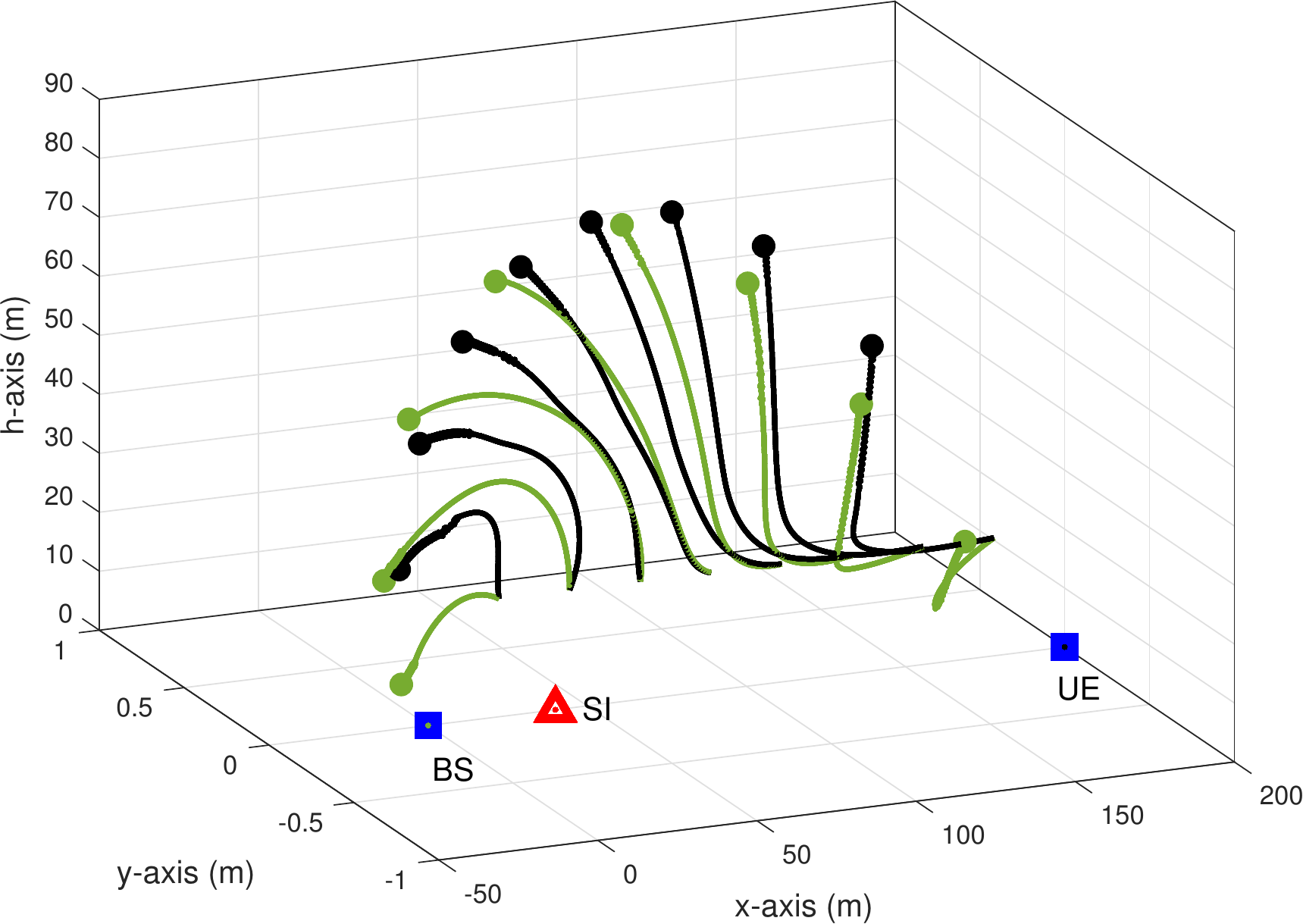}
	\caption{Different trajectories for weighted and unweighted Cheeger constant. Green and black trajectories are based on weighted and regular Cheeger constants, respectively.}
	\label{cheegtraj}
\end{figure}

In Fig.~\ref{fig:3d} and Fig.~\ref{fig:top}, the 3D trajectory of the UAVs are shown in 3D and top views, respectively, assuming 12 UAVs, 2 interferers, and a UE on the ground. We observe that the relaying UAVs adjust their locations in 3D space so as to evade from the interferers, and therefore  improve the desired data flow between the BS and the UE.

\subsubsection{Impact of Weighted Cheeger Constant}
We now look into the impact of using weighted version of Cheeger constant on the desired data flow and 3D trajectories. In Fig.~\ref{cheegflow}, we depict the data flow for the conventional Cheeger constant and weighted Cheeger constant along with assuming 8 UAVs and a single SI. We observe that the weighted version is significantly superior to the conventional one with almost two times the desired data flow. We also exhibit the 3D UAV trajectories in Fig.~\ref{cheegtraj}. Interestingly, weighted Cheeger constant results in 3D trajectories ending up with final UAV locations  closer to both the BS and UE. This may possibly be due to the fact that the bottleneck of the network flow occurs along with the closer links to both BS and UE considering the close proximity of the interferer located on the ground. The weighted Cheeger constant therefore adjusts the final UAV locations so as to make them as close to the BS and UE as possible.

\subsubsection{Interference Avoidance Capability}
\begin{figure}[!t]
	\centering
	\includegraphics[width=0.52\textwidth]{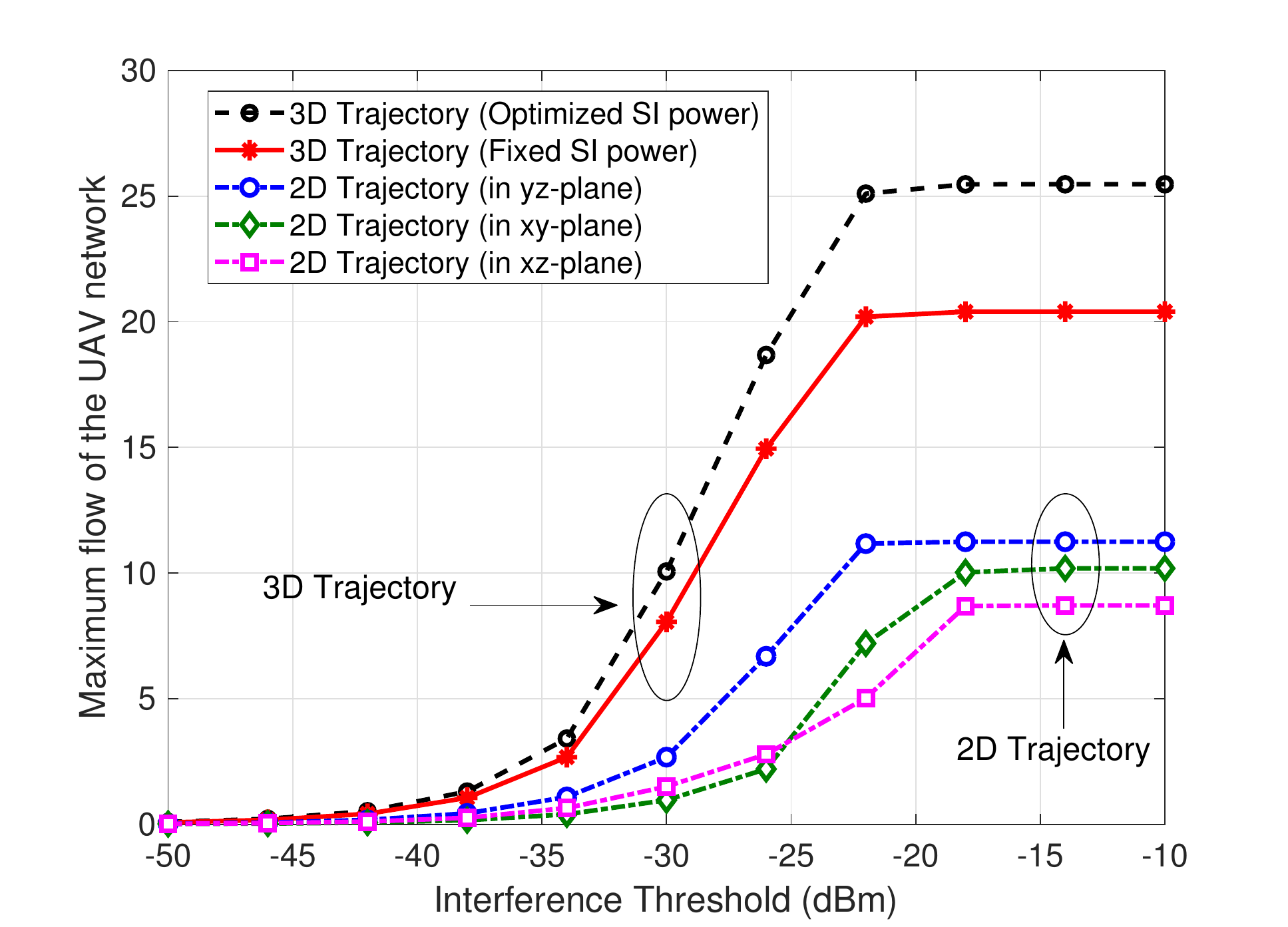}
	\caption{Maximum flow versus interference threshold for 2D and 3D trajectory optimization strategies along with optimal power allocation.}
	\label{interf}
\end{figure}

In Fig.~\ref{interf}, we depict the maximum flow against the interference threshold $\textrm{I}^\textrm{th}$ for the UE altitude of $h^\textrm{UE} \,{=}\,25\,\text{m}$, which may well represent a low-flying UAV as the desired UE. We consider 8 UAVs and one interferer on the ground. We observe that when the relaying UAVs are allowed to optimize their trajectories in 2D only (i.e., in $xy-$, $xz-$, or $yz-$ planes), their performances are always inferior to that of the 3D trajectory optimization.  It can be seen that if the interferer transmit power is optimized, the UAV network performance can be further improved as the interferer will not produce stronger interference beyond satisfying the QoS constraint.

\subsubsection{Impact of the UE Altitude}

\begin{figure}[!t]
	\centering
	\includegraphics[width=0.52\textwidth]{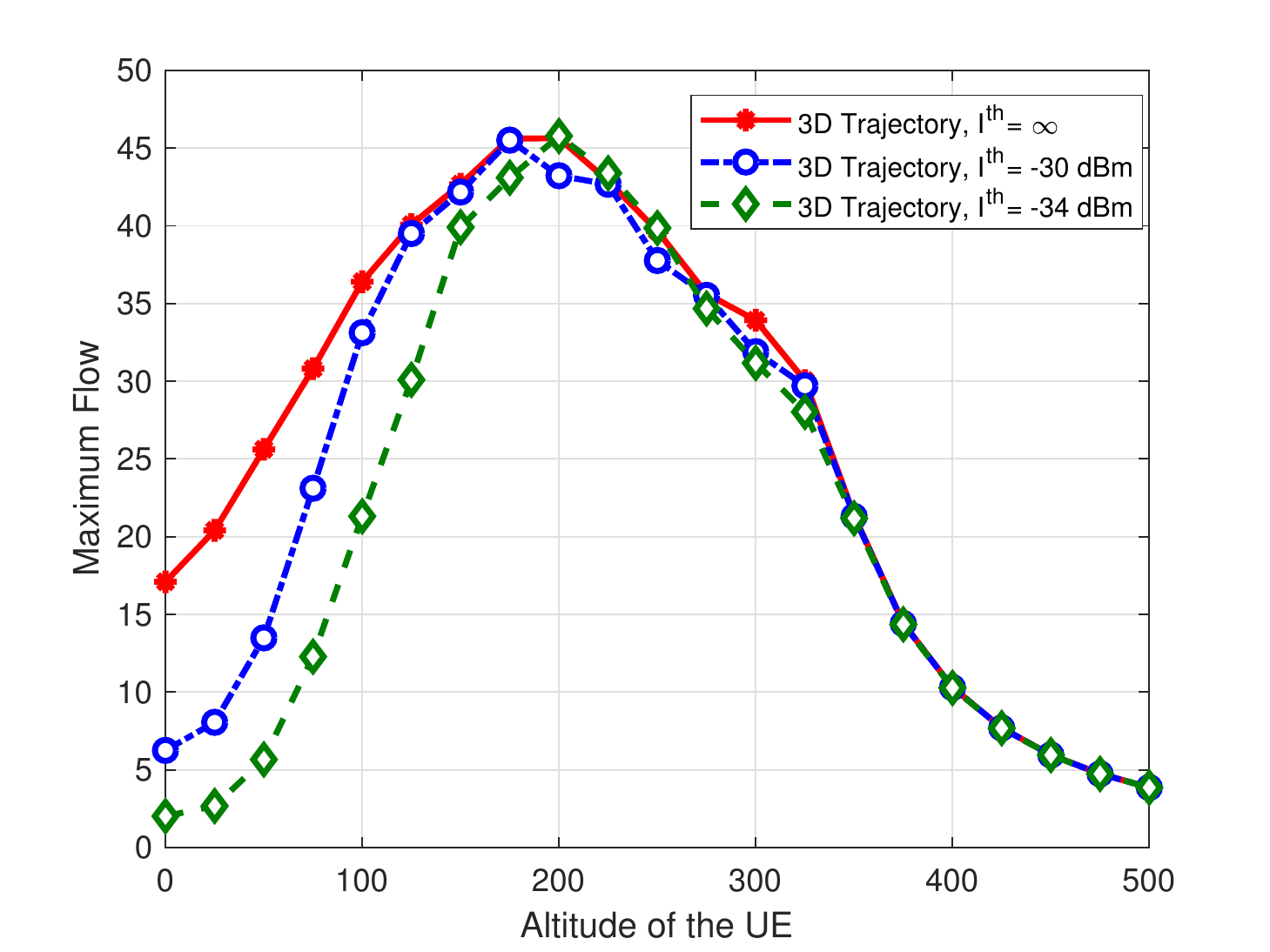}
	\caption{Maximum flow versus UE altitude for interference threshold of $\{{-}34,{-}30\}\,\text{dBm}$ as well as no interference (i.e., $\infty$) with 3D trajectories.}
	\label{UEal}
\end{figure}

\begin{figure}[!t]
	\centering
	\includegraphics[width=0.48\textwidth]{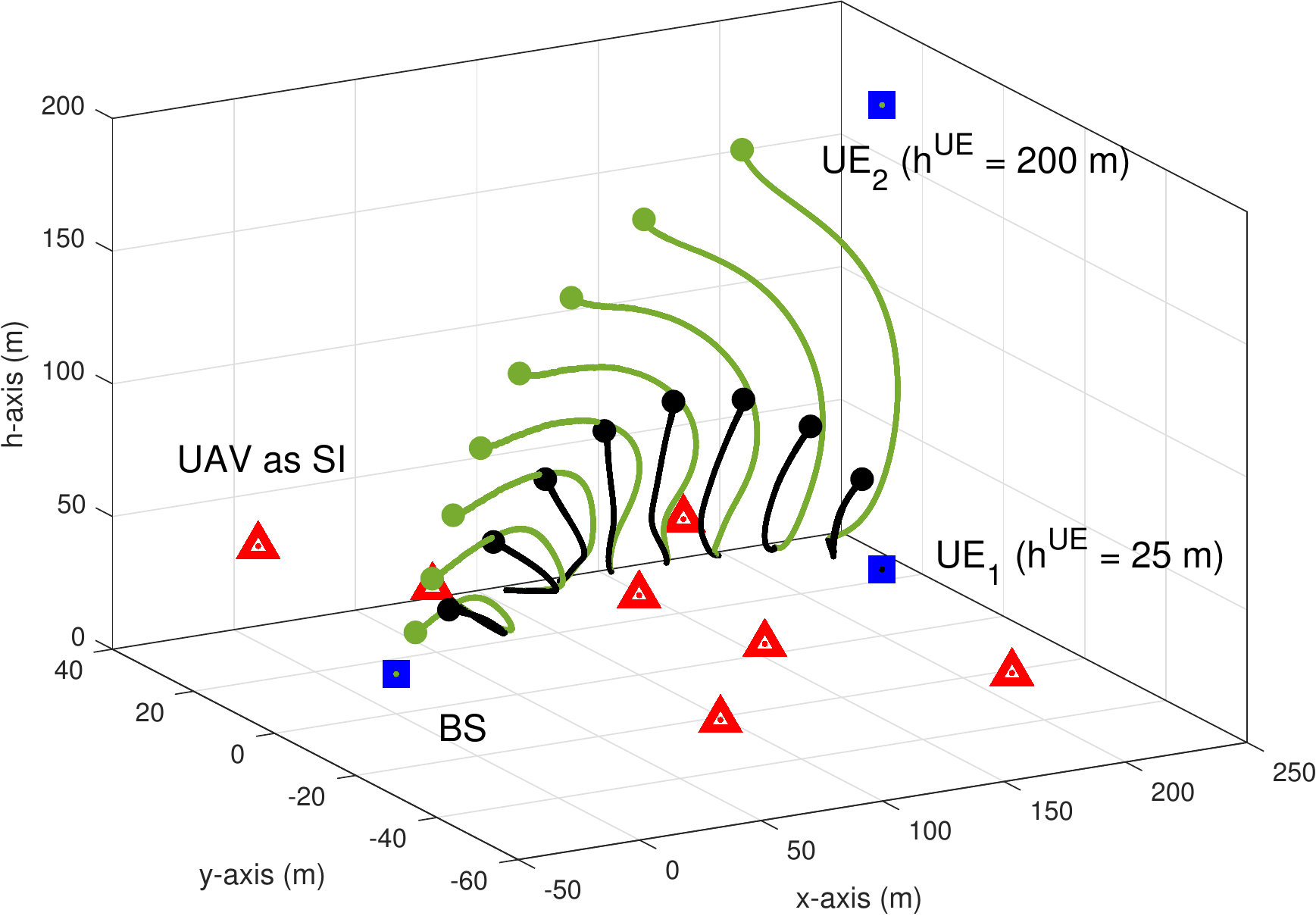}
	\caption{3D trajectories for the results of Fig.~\ref{UEal} at UE altitudes of $\{25,200\}\,\text{m}$ (black trajectory is for 25 m and green trajectory is for 200 m.)}
	\label{twotraj}
\end{figure}

In Fig.~\ref{UEal}, we present the maximum flow along with varying UE altitude of $h^\textrm{UE} \,{\in}\, [0,500]\,\text{m}$, which covers both on-ground and flying UEs. Interestingly, maximum flow improves with increasing altitude till $h^\textrm{UE} \,{=}\, 200\,\text{m}$.  To illustrate this situation, we depict the  3D trajectories of all the 8 UAVs in Fig.~\ref{twotraj} for the UE altitudes of $h^\textrm{UE} \,{=}\, \{25,200\}\,\text{m}$. Moreover, the decrease in the maximum flow after $h^\textrm{UE} \,{=}\, 200\,\text{m}$ is basically due to the increasing path loss, which now becomes more dominant over the interference (even though the interference is also decreasing due to the increasing distance).  


\subsection{Smart Interferer}

\begin{figure}[!t]
	\centering
	\includegraphics[width=0.48\textwidth]{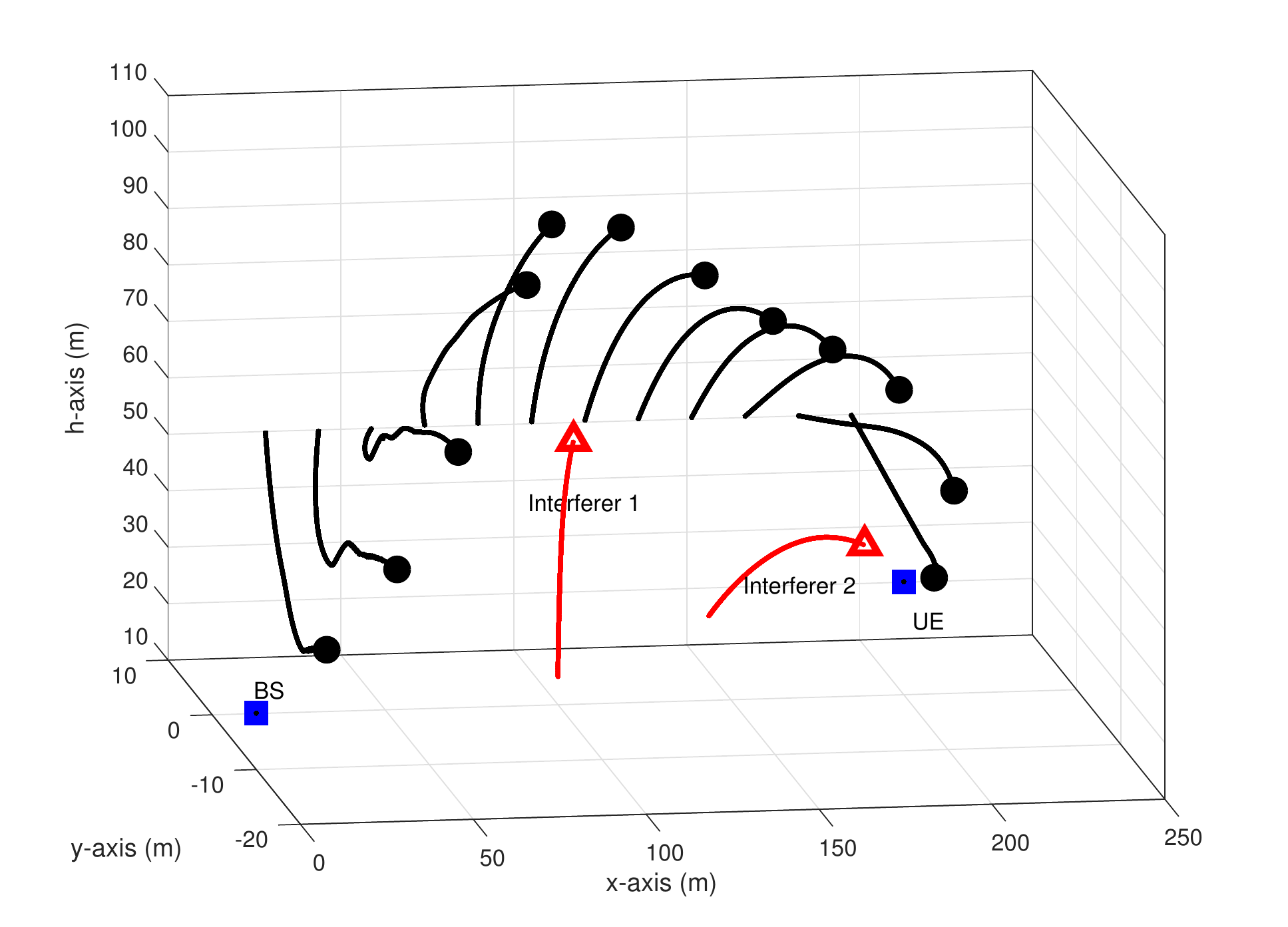}
	\caption{3D trajectories of relaying UAVs, and the smart interferers acting as  UAVs.}
	\label{smart3D}
\end{figure}

\begin{figure}[!t]
	\centering
	\includegraphics[width=0.48\textwidth]{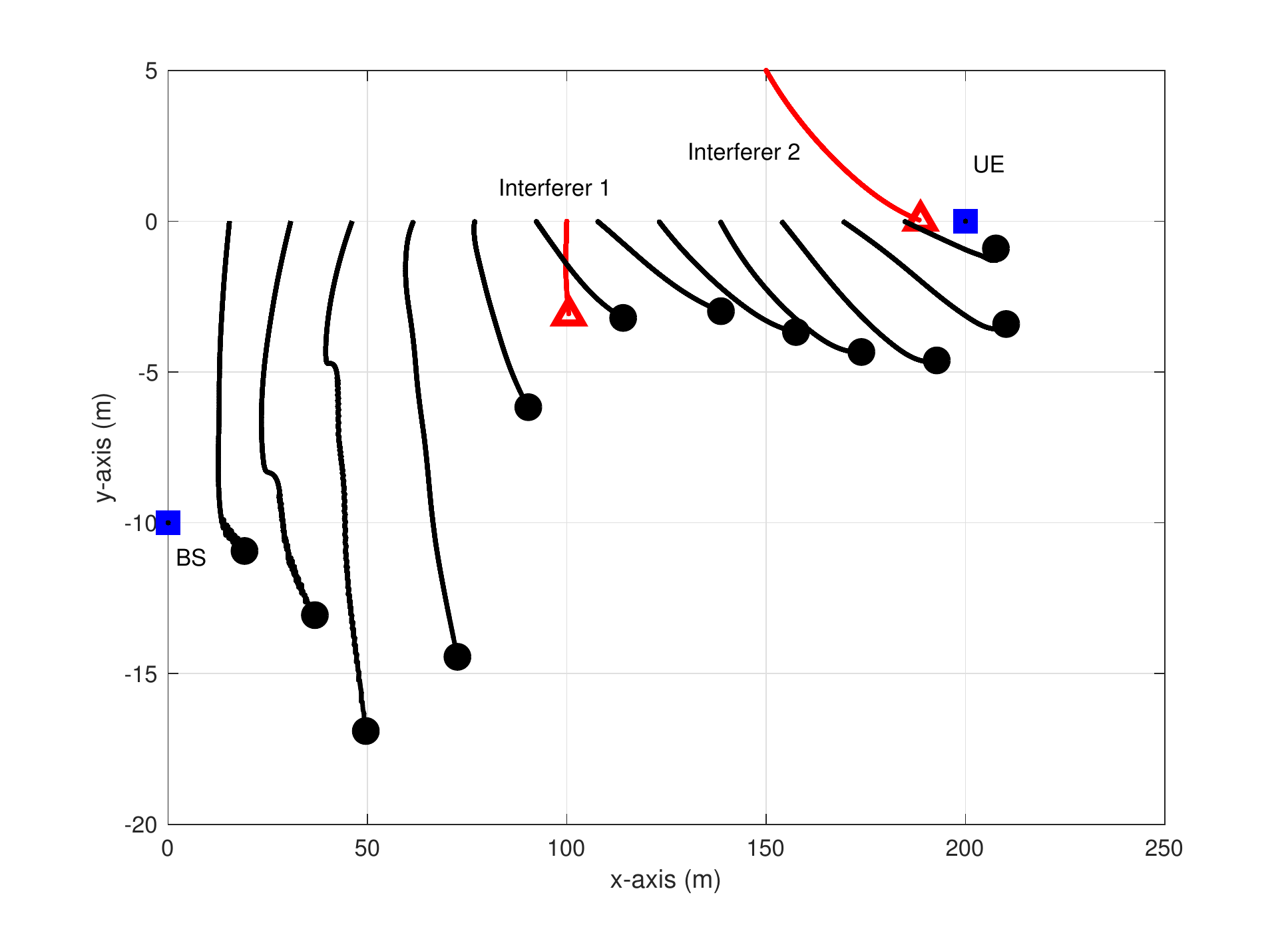}
	\caption{3D trajectories for the UAVs trajectories moving while smart interferers as  UAVs chasing them (top view).}
	\label{smarttop}
\end{figure}

We finally consider a scenario involving two smart interferers, which are basically  malicious UAVs, as discussed in Section IV-B. In order to evaluate its  performance, we assume 12 UAVs and a UE at a height of 40 m (i.e., another UAV). The smart interferers starts its movement at $(100\,\text{m},0, 20\,\text{m})$, and tries to impair the data flow of the UAV-assisted network as much as possible. This may equivalently be viewed as chasing the relaying UAVs. In Fig.~\ref{smart3D} and Fig.~\ref{smarttop}, we depict the UAVs trajectories in 3D and 2D top view, respectively. We observe that while smart interferers are chasing the UAVs to decrease the data flow, the relaying UAVs adjust their locations to get away from the smart interferers.

\begin{figure}[!t]
	\centering
	\includegraphics[width=0.43\textwidth]{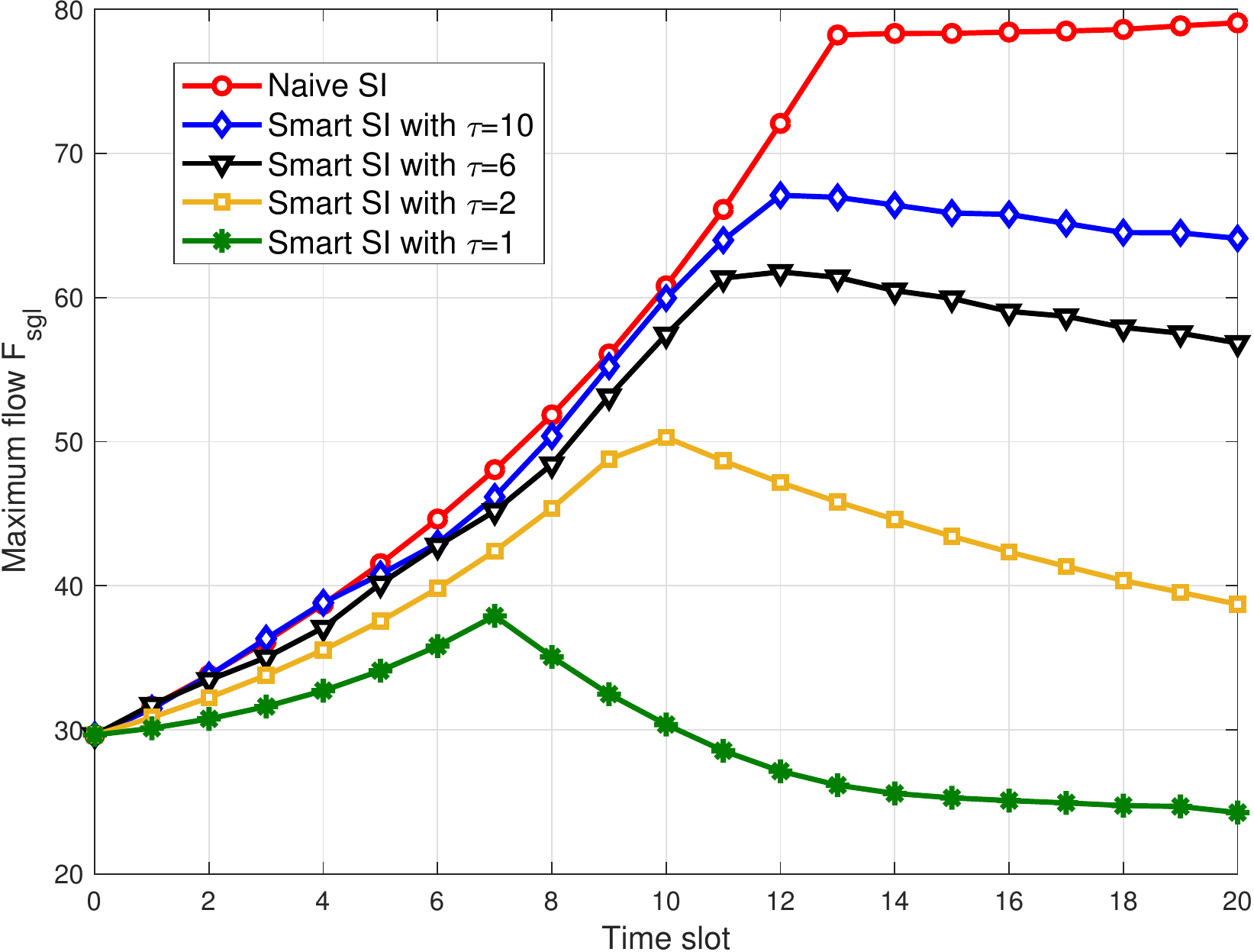}
	\caption{Maximum data flow of the UAV-assisted network with smart interferer, where decreasing $\tau$ corresponds to smarter interferer.}
	\label{flowall}
\end{figure}

Note that we control the smartness of the interferer with the parameter $\tau$, which denotes the ability of the interferer to adjust its 3D location. The value of $\tau \,{=}\, 1$ implies that the interferer can find its direction towards the ``best" 3D location at the same speed of a relaying UAV. Considering a certain amount of time that the smart interferer needs to estimate the UAV-relaying parameters so as to decide the best strategy, we assume $\tau \,{>}\, 1$ for more realistic situations. We  assume $\tau \,{=}\, 2$ for the 3D trajectories presented in Fig.~\ref{smart3D} and Fig.~\ref{smarttop}. We also depict the data flow performance of the UAV-assisted network in Fig.~\ref{flowall} for smart interferers with $\tau \,{\in}\, \{1,2,6,10\}$ and the naive interferers which are not adjusting their locations. We observe that the data flow performance generally degrades as  smart interferers become more capable in adjusting their locations (i.e., smaller $\tau$).

\section{Conclusion}\label{sec:conclusion}
In this paper, we have considered the joint power and 3D trajectory design for a   UAV-assisted relay network in the presence of an already existing network. A joint optimization solution for 3D trajectory design and power allocation is proposed based on spectral graph theory and convex optimization. Moreover, we considered the problem for both reckless and smart interferers. Simulation results show the effectiveness of the proposed algorithm in improving the maximum flow and interference mitigation. In particular, we have shown that the proposed 3D trajectory design can increase the UAV network maximum flow by more than two times while the interference threshold is satisfied on the primary existing network. Moreover, we have shown that there exists an optimal altitude for the UE as a UAV that maximizes the maximum flow of the UAV network. On the other hand, we observed that the UAVs can reconfigure their locations to evade the smart  interferer, while  smart interferers chase the UAVs so as to decrease the maximum flow of the network by increasing the interference resulted from decreasing the path loss effect.  

\bibliographystyle{IEEEtran}
\bibliography{IEEEabrv,references}

\end{document}